\documentclass{aastex631}
\usepackage{multirow}
\usepackage{scalerel}
\usepackage{amsmath}	% Advanced maths commands

\begin{document}

\title{A comprehensive kinematic model of the LMC disk from star clusters and field stars 
using \\ \textit{Gaia} DR3: Tracing the disk characteristics, rotation, bar, and the outliers}
\shorttitle{Kinematics of LMC using star clusters}
\shortauthors{Dhanush et al.}

\correspondingauthor{S. R. Dhanush}
\email{srdhanushsr@gmail.com}
\author[0009-0007-0388-3143]{S. R. Dhanush}
\affiliation{Indian Institute of Astrophysics, Bangalore, 560034,  India}
\affiliation{Pondicherry University, R.V. Nagar, Kalapet, 605014, Puducherry, India}
\author[0000-0003-4612-620X]{A. Subramaniam}
\affiliation{Indian Institute of Astrophysics, Bangalore, 560034,  India}
\author[0000-0002-5331-6098]{S. Subramanian}
\affiliation{Indian Institute of Astrophysics, Bangalore, 560034,  India}

\begin{abstract}
The internal kinematics of the Large Magellanic Cloud (LMC) disk have been modeled by several studies using different tracers with varying coverage, resulting in a range of parameters. Here, we modeled the LMC disk using 1705 star clusters and field stars, based on a robust Markov Chain Monte Carlo (MCMC) method, using the \textit{Gaia} DR3 data. The dependency of model parameters on the age, coverage, and strength of the clusters are also presented. This is the first comprehensive 2D kinematic study using star clusters. Red clump (RC) stars and young main-sequence stars are also modeled for comparison. The clusters and field stars are found to have distinctly different kinematic centers, disk inclination, position angle of the line of nodes, and scale radius. We also note a significant radial variation of the disk parameters. Clusters and young stars are found to have a large residual proper motion and a relatively large velocity dispersion when compared to the RC field population, which could be due to perturbation from the bar and spiral arms. We traced the presence of large residual proper motion and non-circular motion among clusters likely to be due to the bar and detected a decrease in the scale radius as a result of the possible evolution of the bar. The kinematically deviant clusters point to a spatio-temporal disturbance in the LMC disk, matching with the expected impact factor and time of the recent collision between the LMC and the Small Magellanic Cloud. 

\end{abstract}



\section{Introduction} \label{sec:intro}
The Magellanic Clouds (MCs) consist of two irregular dwarf galaxies observable from the Southern Hemisphere. The Large Magellanic Cloud (LMC), the larger of the pair, is situated at 49.59$\pm$0.09 kpc away (\citealt[hereafter, P20]{lmc_dit2019Natur.567..200P}), while its smaller companion, the Small Magellanic Cloud (SMC), lies at a distance of 62.44$\pm$0.47 kpc \citep{smc_dist2020ApJ...904...13G}. The close proximity of the MCs to the Milky Way (MW) offers a unique opportunity to study them and gain valuable insights into the mechanisms that drive galaxy mergers and their consequences for galaxy evolution.

The LMC-SMC-MW is an interacting system \citep{put1998Natur.394..752P,Diaz_2012,Hammer_2015}. Traditionally, the kinematics of the LMC was thought to be modified by its interactions with the MW. In the last couple of decades, it was believed that the LMC was experiencing its first encounter with the MW, as indicated in the studies by  \cite{kal22006ApJ...652.1213K,kal2006ApJ...638..772K}, and \cite{beslakaliay2007ApJ...668..949B}. However, the recent study by \cite{vas10.1093/mnras/stad2612} indicates the potential scenario where the MCs are in their second passage around the MW. Either way, the current morphology of the LMC is predominantly influenced by the interactions with the neighboring SMC rather than the MW (\citealt[hereafter B12]{bes2012MNRAS.421.2109B}; \citealt{yozi_bekki_lmc_smc2014MNRAS.439.1948Y}; \citealt{choi2018aApJ...866...90C,Choi_2018}). 

There have been significant efforts to understand the internal kinematics of the LMC. They were previously studied primarily using the H I gas and stars in the LMC \citep{kim1998ApJ...503..674K,van2002AJ....124.2639V,olsen2007ApJ...656L..61O,Indu2015A&A...573A.136I}. Recently, \citet[hereafter W20]{wan2020MNRAS.492..782W} used the all-sky data release of SkyMapper (DR1) to investigate the kinematics of the LMC across various stellar populations, including carbon stars, RGB stars, and young stars. Also, the kinematics of central regions of the LMC is studied by \citet[hereafter N22]{florian2022MNRAS.512.5423N} using the Visible and Infrared Survey Telescope for Astronomy (VISTA) survey of the Magellanic Clouds
system (VMC; \citealt{cioni2011A&A...527A.116C}). Recent studies with \textit{Gaia} Early Data Release 3 (EDR3) and Data Release 3 (DR3), revealed the structure and velocity maps of the LMC (\citealt[hereafter G21]{gaia2021A&A...649A...7G}; \citealt[hereafter J23]{Jimenez2023A&A...669A..91J}). Also, \citet[hereafter C22]{choi2022ApJ...927..153C} performed the 3D kinematic model of the LMC, with the red clump population using \textit{Gaia} EDR3. Their findings suggest a direct LMC-SMC collision in the last 250 Myr.

Star clusters are excellent tracers of star formation history. So far, star clusters are used to trace the cluster formation (CF) peaks (\citealt{glatt2010}; \citealt{nayak2016MNRAS.463.1446N}; \citealt{SRD10.1093/mnras/stae096}, hereafter D24), chemical enrichment of the disk \citep{hill2000A&A...364L..19H,Johnson_2006,palma201510.1093/mnras/stv762,chillingr2018ApJ...858...63C,narloch2022A&A...666A..80N}  and in a few cases, kinematics of the LMC disk using radial velocity estimations \citep{schommer,Bennet_2022}. The availability of the proper motion (PM) data from the \textit{Gaia} DR3 catalog opens up a new method to explore the LMC disk kinematics using star clusters. Recently D24 parameterised 1710 star clusters in the LMC using the \textit{Gaia} DR3 data. The PM of the cluster estimated using the median PM value of the members therefore presents a golden data set to explore the LMC disk kinematics using tangential velocity. As the ages of these clusters are already estimated, the data set will allow us to explore the dependency of kinematic properties on the age of the cluster. D24 also identified a nearby field region for all the 1710 clusters, and this population can be used as a control/ comparison population. This study therefore explores the LMC disk kinematics using the unique data set of D24.

The aim of this study is to model the LMC disk and estimate the kinematic parameters using a robust Markov Chain Monte Carlo (MCMC) method. The data set allows us to explore their dependency on various factors such as age, spatial coverage of the population as well as the richness of clusters. The kinematic parameters of the LMC disk are derived for the first time using star clusters in this study. We also quantify the residual PM after fitting the model and trace the kinematic outliers among clusters. 

The paper is arranged as follows. In Section \ref{sec_2}, we describe the data sets used for our modeling. In Section \ref{sec_3}, we describe the procedures of our kinematic model. In Section \ref{sec_4}, we present the estimated kinematic parameters of our data sets, followed by the estimation of the residual PM. The discussions based on our results are presented in Section \ref{sec_5}, followed by Section \ref{sec_6}, which summarizes our work.

\section{Data}\label{sec_2}
The data sets of our modeling were derived from D24, which involved parameterizing star clusters in MCs with \textit{Gaia} DR3 data to estimate their age, metallicity, distance modulus, and extinction. The study parameterized 1990 star clusters within the MCs, among which 1710 were in the LMC and 280 were in the SMC. These clusters underwent thorough field star subtraction using nearby regions, as explained in Section 2 of D24. For the 1710 LMC clusters, we used the cleaned cluster regions and their associated field regions as data sets to perform the kinematic model. First, we calculated the median of the PM and its standard error for stars within each cluster and their associated field region in both the RA and DEC directions. Then, we applied a PM cutoff of $<$ (median PM$+$3$\sigma$) based on the PM distribution of field regions. We used it for clusters and field regions both to eliminate a few outliers in the galaxy that show relatively large PM. We retained 1705 clusters after this selection. The primary modeling was performed using the clusters and their associated field regions following this selection process. We further performed kinematic modeling with data sets based on the control population, cluster richness,  cluster age, and spatial coverage of the LMC. The details of these data sets are enumerated below. 

\begin{enumerate}
    \item The control population for the model is obtained from the color-magnitude-diagrams (CMDs) of field regions associated with each cluster. The choice of control population was made to understand the kinematic nature of the cluster and field population. We chose the young main-sequence (MS) stars within the window of \textit{$G$} $<$ 19.5 mag and (\textit{$G_{BP}$}-\textit{$G_{RP}$}) $<$ 0.5, then red clump (RC) stars based on the selection criteria from CMD of the LMC sources as mentioned in \cite{saroo2022A&A...666A.103S}. Several field regions associated with clusters do not have significant young MS stars in the CMD. We put a cut-off of young MS stars $\geq$5 to further select the samples for the control population. We are left with 1484 candidates in the four categories for modeling, the clusters, nearby field regions, young MS stars, and RC stars. Out of this final selection, we note that the $\sim$ 75\% of the field regions considered have young MS stars and RC stars with a strength of 20 and 50 stars, respectively.
    \item We note that the average number of members in each cluster in our study is $\sim$ 40. To investigate potential biases in the data sets, we conducted modeling of groups based on the cluster members' strength across a range from 10 to 60 members, with intervals of 10. The field regions associated with these richness groups were also modeled to compare with the nature of clusters. We created 12 models for the cluster groups and field groups based on cluster richness.
    \item To understand the age dependency of the kinematic parameters, we performed modeling with respect to clusters spanning from older to younger age groups. We used age groups in ranges of $\log(t)$, Age group-1: $[\geq9.10,<9.55]$, Age group-2: $[\geq8.65,<9.10]$, Age group-3: $[\geq8.0,<8.65]$, and Age group-4: $[\geq6.55,<8.0]$  (as defined in D24). The associated nearby field regions were also modeled with respect to each age group. It is to be noted that the age-wise grouping is only valid for the clusters, and the associated field regions are age-wise heterogeneous but analyzed for comparison.
    \item We investigated the variations in kinematic parameters concerning the spatial extent of the LMC. Employing circular regions, we analyzed clusters and associated field regions, ranging from 2$^\circ$ to 7$^\circ$ in increments of 0.5$^\circ$ from the kinematic center on the sky plane. We used the kinematic center that we calculated from the primary model encompassing clusters and field stars for the radially separated groups.
\end{enumerate}

Table \ref{tab1} summarizes the data sets that are modeled. The subsequent section covers the implementation of the kinematic model using the above data sets.

\begin{table}
    \centering
    \caption{The data sets used in the kinematic model of the LMC are summarized here. The length ($L_d$) and the number ($N_d$) of data sets are provided as well. Clusters and associated field regions are labeled based on the cluster richness in the 3$^{rd}$ row, such as C$_{10}$ representing clusters with 10 or more stars, and so on. A total of 48 data sets are used in our study to perform the kinematic modeling.}
    \label{tab1}
    \renewcommand{\arraystretch}{1.35} % Adjust the value as needed
    \begin{tabular}{lcc}
        \hline\hline
        \textbf{Data set} & \textbf{L$_d$} & \textbf{N$_d$} \\
        \hline
        Clusters, Nearby field regions & 1705 & 2 \\
        (Primary sample) & & \\
        \hline
        \multirow{1}{*}{Clusters, Nearby field regions, } & \multirow{1}{*}{1484} & \multirow{1}{*}{4} \\
         \multirow{1}{*}{Young MS stars, RC stars} & \multirow{1}{*}{} & \multirow{1}{*}{}\\
           \multirow{1}{*}{(Control sample)} & \multirow{1}{*}{} & \multirow{1}{*}{}\\
        \hline
        \multirow{1}{*}{C$_{10,..,60}$, F$_{10,..,60}$ } & \multirow{1}{*}{1581, 973, 643,} & \multirow{1}{*}{12} \\
         \multirow{1}{*}{(groups based on cluster richness)} & \multirow{1}{*}{455, 334, 257} & \multirow{1}{*}{}\\
         \hline
         \multirow{1}{*}{Cluster age groups, } & \multirow{1}{*}{316, 579, 551, 260} & \multirow{1}{*}{8} \\
         \multirow{1}{*}{Nearby field regions} & \multirow{1}{*}{} & \multirow{1}{*}{}\\
         \hline
          \multirow{1}{*}{Clusters, Nearby field regions } & \multirow{1}{*}{372, 560, 551, 768,} & \multirow{1}{*}{22} \\
          \multirow{1}{*}{(groups based on spatial coverage,} & \multirow{1}{*}{958, 1104, 1263, 1401,} & \multirow{1}{*}{}\\
         \multirow{1}{*}{radii: 2$^{\circ}$ to 7$^{\circ}$, step size: 0.5$^{\circ}$)} & \multirow{1}{*}{1604, 1667, 1688 } & \multirow{1}{*}{}\\
         \hline\hline
    \end{tabular}
\end{table}

\section{Kinematic model of the LMC}\label{sec_3}

We performed the kinematic model of the LMC corresponding to the observed median PM for the data sets referenced in Section \ref{sec_2}. The following subsections provide a detailed explanation of the theoretical background we adopted in our modeling and the Bayesian methodology employed to estimate the optimal kinematic parameters for the data sets.

\subsection{Analytical background of the modeling}
The model PM for the candidates in the data sets is formulated based on the equations outlined by \citet[hereafter, V02]{van2002AJ....124.2639V}. These equations define the directional vectors of local PM in the West and North (M$_W$, M$_N$) in the sky plane using a series of 12 independent model parameters. The parameters selected for our model encompass the inclination of the LMC disc ($i$), the position angle of the line of nodes measured from West ($\theta$), dynamic centers ($\alpha_0$, $\delta_0$), the amplitude of the tangential velocity of the LMC's center of mass ($v_t$), the tangential angle made by $v_t$ ($\theta_t$), scale radius ($R_0$), optimization factor ($\eta$), and the amplitude of the rotational velocity ($v_0$). Our modeling process is aimed at determining the most fitting values for these nine kinematic parameters. It's important to note that our modeling was confined to the LMC's sky plane due to the absence of line-of-sight velocity components in our data sets.

The velocity of sources at any given point ($\alpha$, $\delta$) within the LMC sky plane is primarily composed of three components. It consists of the velocity of the center of mass (COM) of the LMC, the velocity component associated with the precession and nutation of the galaxy, and the internal rotation component of the galaxy. We assumed that there is no precession and nutation of the LMC disc within the spatial coverage of the galaxy (less than 7$^\circ$) considered in this study (V02, N22).
We assumed the RA and DEC of the field regions associated with each cluster to have the same spatial coordinates as the cluster centers. Also, we assumed a fixed distance of 49.59 kpc (P20) to the LMC center, D$_0$ and a line-of-sight velocity of the COM, $v_{sys}$, as 262.2 km s$^{-1}$ (V02). The parametric form for the rotational velocity in the disc plane was adopted from similar kinematic models used in previous studies of the LMC (G21, N22).

The following subsection deals with the method and estimation of the best-fitting kinematic parameters for the data sets summarized in Table \ref{tab1}.

\subsection{Modeling procedure}
We found the best-fitting 48 models and the associated kinematic parameters for 48 data sets referenced in Table \ref{tab1}. We performed the fitting of the parameters using the MCMC serial stretch move sampling algorithm introduced by \citealt{gdman2010CAMCS...5...65G}. The code is developed and implemented in C language. A similar MCMC approach is used in the studies by W20, C22 and N22. The observed PM for the sources in data sets are along the RA and DEC directions, but the model formulation is for the PM in the West and North directions. For that, we took the negative of the source median PM in the RA direction. Now the model PM (M$_{W,m}$, M$_{N,m}$) and the observed PM (M$_{W,o}$, M$_{N,o}$) can be used to construct the log-likelihood function ($\ln\mathcal{L}$), which can be used with the associated West and North direction standard errors of observed data sets ($\sigma_{W,o}$,$\sigma_{N,o}$) to sample the best fitting parameter with MCMC. The equation for $\ln\mathcal{L}$ is given by,
\begin{align}
    \ln\mathcal{L} &= -0.5 \sum_{i=1}^{n} \left[ \ln \left( 2\pi\sigma_{W,o,i}^2 \right) + \frac{ \left( \mu_{W,o,i} - \mu_{W,m,i} \right)^2 }{\sigma_{W,o,i}^2} \right. \nonumber \\
    &\quad + \ln \left( 2\pi\sigma_{N,o,i}^2 \right) + \left. \frac{ \left( \mu_{N,o,i} - \mu_{N,m,i} \right)^2 }{\sigma_{N,o,i}^2} \right]
\end{align}

The priors for the model were uniformly chosen, except for the rotation velocity amplitude, $v_0$, for which we used a Gaussian prior of 76 km s$^{-1}$ (\citealt{vanderMarel_2014}; G21; N22). When working with modeling datasets based on cluster age groups, we note that the spatial distribution of clusters in the sky plane is not homogeneous. This results in the non-convergence of $\eta$ in its posterior sampling distributions while performing the MCMC with a uniform prior. To avoid this, we employed Gaussian weighting on $\eta$ specifically for these age-based datasets, using the corresponding value of the parameter estimated from the primary model encompassing all cluster and field regions.

We executed 2000 steps of MCMC iteration involving 200 walkers evolving sequentially at each step. From the sampled posterior values for the nine parameters, we focused on the final 50\%, calculating their median alongside the 16$^{th}$ and 84$^{th}$ percentile errors for estimation. 
In the following section, we present the comprehensive results obtained from the above modeling procedures.

\section{Results}\label{sec_4}

\begin{figure*}
    \centering
       \includegraphics[width=1\linewidth]{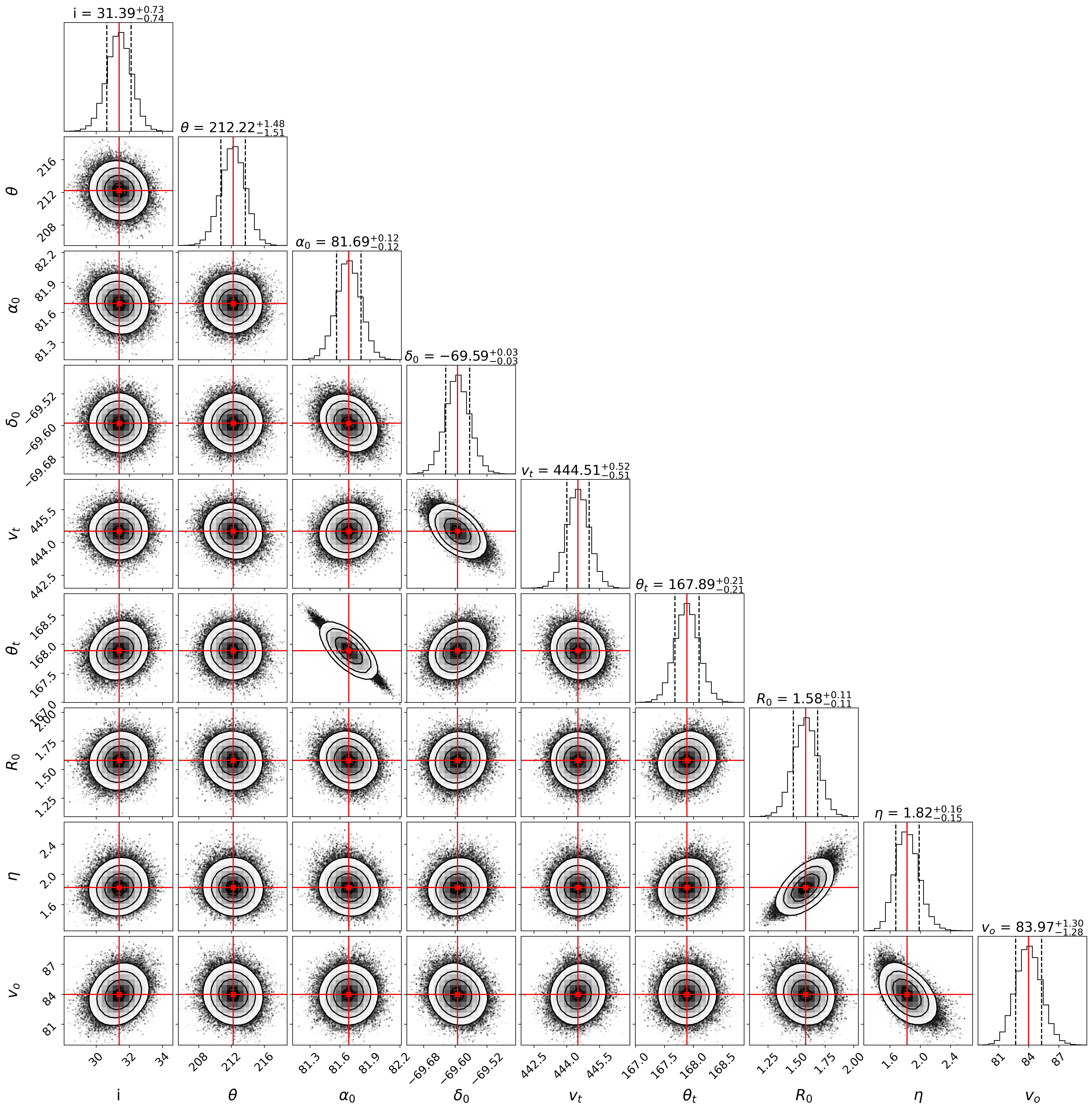}
       \caption{The corner plot representing the sampled posterior distribution of nine kinematic parameters for the primary cluster data set is shown here. The vertical red lines represent the median values, and the black dashed lines represent the 16th and 84th percentiles.}
       \label{fig:mcmc_cluster_full} 
\end{figure*}

\begin{figure*}
    \centering
       \includegraphics[width=1\linewidth]{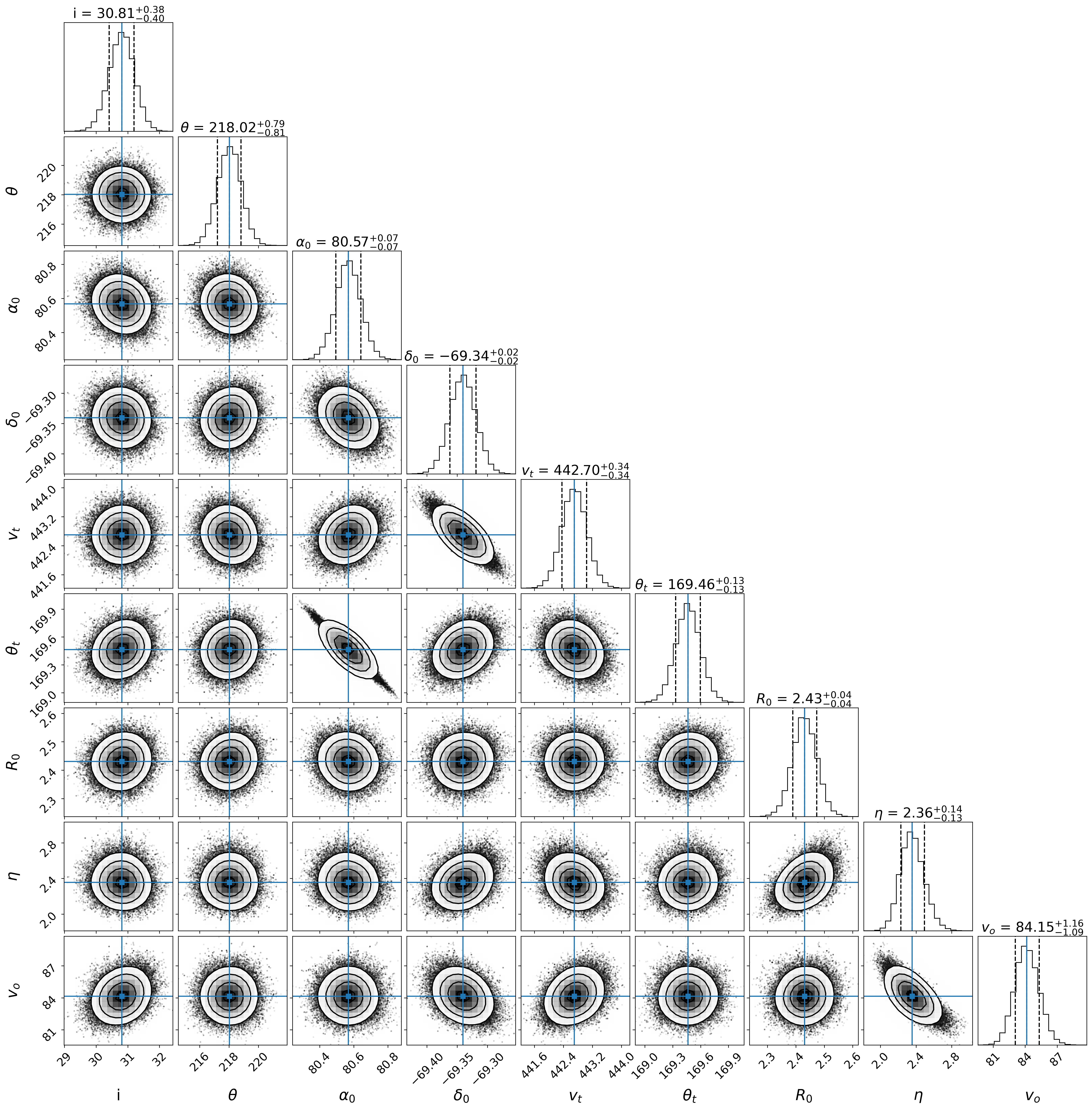}
       \caption{The corner plot representing the sampled posterior distribution of nine kinematic parameters for the primary field data set is shown here. The vertical blue lines represent the median values, and the black dashed lines represent the 16th and 84th percentiles.}
       \label{fig:mcmc_field_full} 
\end{figure*}

This is the first 2D kinematic model of the LMC employing clusters and neighboring field regions with \textit{Gaia} DR3 data. The data coverage of the LMC considered in this study is within $\sim$ 7$^\circ$ from the LMC center. The primary modeling is performed with 1705 clusters and nearby field regions, estimating nine best-fitting model parameters defining the kinematic properties of the LMC. Figures \ref{fig:mcmc_cluster_full} and \ref{fig:mcmc_field_full} show the sampled posterior distribution of the parameters for clusters and nearby field regions, respectively. We also performed the parameter estimation for the rest of the data sets mentioned in Table \ref{tab1}. 

%%%%%%%%%%%%%%%%%%%%%%%% MCMC figures

%%%%%%%%%%%%%%%%%%%%%%%%%%%%%%

In the following subsections, we provide the estimated kinematic parameters obtained for all the modeled data sets. Additionally, we present the residual PM derived from comparing the model and observed data.

\subsection{Cluster and Field kinematics}\label{clust_field_kineamtic}
\begin{figure*}
    \centering
       \includegraphics[width=1\linewidth]{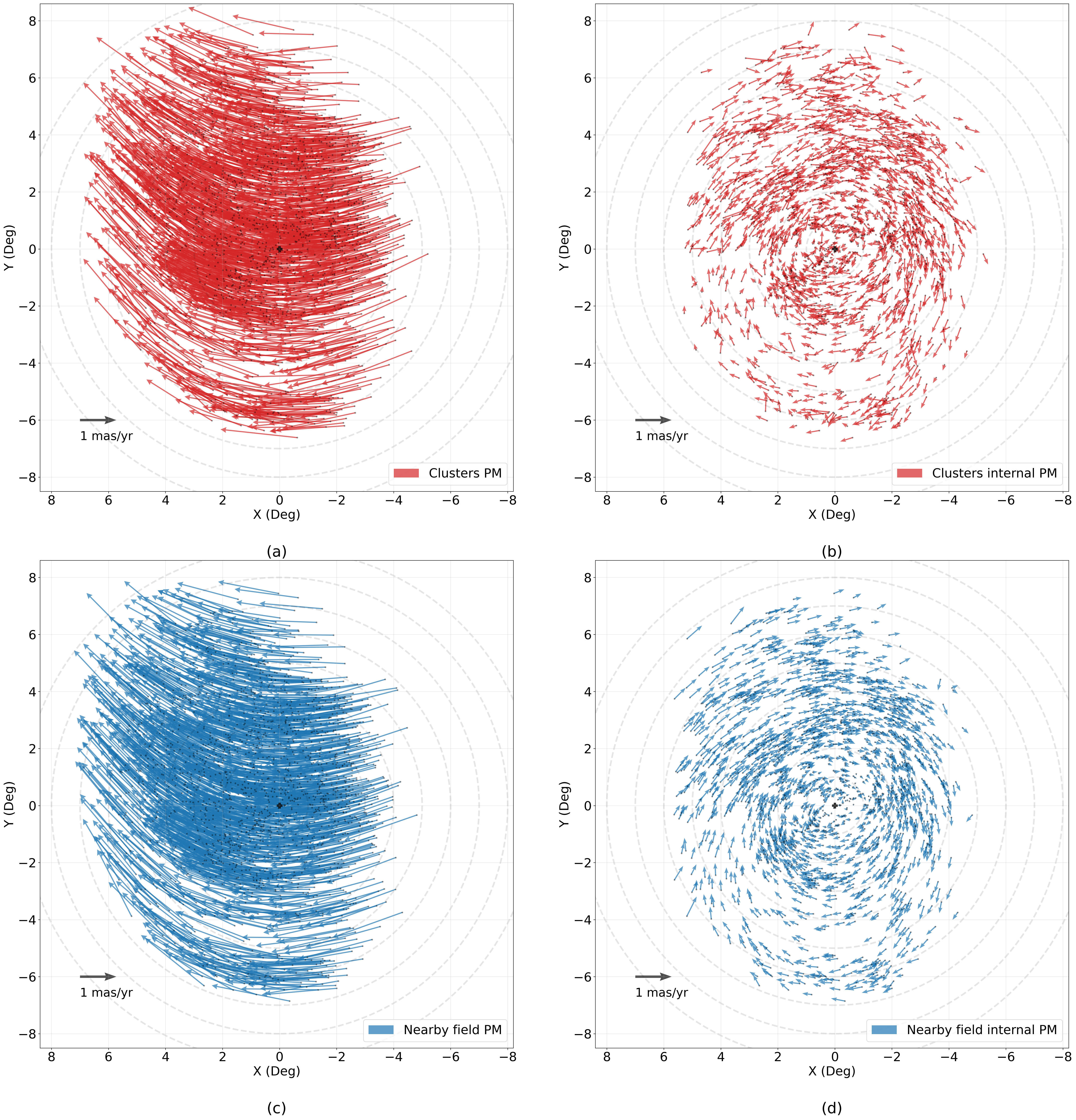}
       \caption{The observed PM plots for the clusters and field in the LMC sky plane are shown here. (a) Observed bulk motion of clusters; (b) Internal rotation PM of clusters; (c) Observed bulk motion of field; (b) Internal rotation PM of field.}
       \label{fig:PM_plots} 
\end{figure*}

Table \ref{tab2_paraemters} provides the estimated kinematic parameters for the primary model involving clusters and field stars. The position angle of the line of nodes is conventionally measured from the North of the galaxy, whereas the modeling we performed involved measuring from the West. Therefore, we define the position angle of the line of nodes measured from North (PA) as $\Theta$, which is ($\theta$-$90^{\circ}$) throughout the sections. The COM PM values in the West and the North directions ($\mu_{W,com}$ and $\mu_{N,com}$) can be obtained using the estimated tangential velocity vector estimated in our models (using  $v_t$ and $\theta_t$ and Equation 10 in V02). The orthographic projections of the sky coordinate and PM vectors are performed using equations 1, 2, and 3, as mentioned in G21. Figure \ref{fig:PM_plots} shows the bulk motion and rotation of the LMC obtained with clusters and field by subtracting the COM PM values estimated from the model. Below we compare the notable kinematic properties of clusters and field regions obtained from the modeling.  
\begin{table}
    \centering

    \caption{The kinematic best-fitting parameters obtained for the LMC from the primary model of clusters and nearby field regions are tabulated below.}
    \label{tab2_paraemters}
\renewcommand{\arraystretch}{1.5}
    \begin{tabular}{l@{\hspace{0.65em}}c@{\hspace{0.65em}}c@{\hspace{0.65em}}c@{\hspace{0.65em}}c@{\hspace{0.65em}}c@{\hspace{0.65em}}c@{\hspace{0.65em}}c@{\hspace{0.65em}}c@{\hspace{0.65em}}c@{\hspace{0.65em}}}
        \hline\hline
        \textbf{Data} & \textbf{$i$} & \textbf{$\Theta$} & \textbf{$\alpha_0$}& \textbf{$\delta_0$}& \textbf{$v_t$}& \textbf{$\theta_t$}& \textbf{$R_0$}& \textbf{$\eta$}& \textbf{$v_0$} \\
         & (deg) & (deg) & (deg) & (deg) & (km s$^{-1}$) & (deg) & (kpc) &  & (km s$^{-1}$) \\
        \hline
       Clusters& 31.39$^{\hspace{0.01cm}\scaleto{+0.73}{4.7pt}}_{\hspace{0.01cm}\scaleto{-0.74}{4.7pt}}$ & 122.22$^{\hspace{0.01cm}\scaleto{+1.48}{4.7pt}}_{\hspace{0.01cm}\scaleto{-1.51}{4.7pt}}$ & 81.69$^{\hspace{0.01cm}\scaleto{+0.12}{4.7pt}}_{\hspace{0.01cm}\scaleto{-0.12}{4.7pt}}$ & -69.59$^{\hspace{0.01cm}\scaleto{+0.03}{4.7pt}}_{\hspace{0.01cm}\scaleto{-0.03}{4.7pt}}$ & 444.51$^{\hspace{0.01cm}\scaleto{+0.52}{4.7pt}}_{\hspace{0.01cm}\scaleto{-0.51}{4.7pt}}$ & 167.89$^{\hspace{0.01cm}\scaleto{+0.21}{4.7pt}}_{\hspace{0.01cm}\scaleto{-0.21}{4.7pt}}$ & 1.58$^{\hspace{0.01cm}\scaleto{+0.11}{4.7pt}}_{\hspace{0.01cm}\scaleto{-0.11}{4.7pt}}$ & 1.82$^{\hspace{0.01cm}\scaleto{+0.16}{4.7pt}}_{\hspace{0.01cm}\scaleto{-0.15}{4.7pt}}$ & 83.97$^{\hspace{0.01cm}\scaleto{+1.3}{4.7pt}}_{\hspace{0.01cm}\scaleto{-1.28}{4.7pt}}$ \\
         \hline
        Nearby field& 30.81$^{\hspace{0.01cm}\scaleto{+0.38}{4.7pt}}_{\hspace{0.01cm}\scaleto{-0.4}{4.7pt}}$ & 128.02$^{\hspace{0.01cm}\scaleto{+0.79}{4.7pt}}_{\hspace{0.01cm}\scaleto{-0.81}{4.7pt}}$ & 80.57$^{\hspace{0.01cm}\scaleto{+0.07}{4.7pt}}_{\hspace{0.01cm}\scaleto{-0.07}{4.7pt}}$ & -69.34$^{\hspace{0.01cm}\scaleto{+0.02}{4.7pt}}_{\hspace{0.01cm}\scaleto{-0.02}{4.7pt}}$ & 442.7$^{\hspace{0.01cm}\scaleto{+0.34}{4.7pt}}_{\hspace{0.01cm}\scaleto{-0.34}{4.7pt}}$ & 169.46$^{\hspace{0.01cm}\scaleto{+0.13}{4.7pt}}_{\hspace{0.01cm}\scaleto{-0.13}{4.7pt}}$ & 2.43$^{\hspace{0.01cm}\scaleto{+0.04}{4.7pt}}_{\hspace{0.01cm}\scaleto{-0.04}{4.7pt}}$ & 2.36$^{\hspace{0.01cm}\scaleto{+0.14}{4.7pt}}_{\hspace{0.01cm}\scaleto{-0.13}{4.7pt}}$ & 84.15$^{\hspace{0.01cm}\scaleto{+1.16}{4.7pt}}_{\hspace{0.01cm}\scaleto{-1.09}{4.7pt}}$ \\
        \hline\hline
    \end{tabular}
\end{table}
\begin{itemize}
  \item  $\mu_{W,com}$ and $\mu_{N,com}$ were estimated as $1.849\pm0.003$ mas yr$^{-1}$ and $0.397\pm0.007$ mas yr$^{-1}$ for the clusters. Similarly, for the field regions, the values are $1.851\pm0.002$ mas yr$^{-1}$ and $0.344\pm0.004$ mas yr$^{-1}$. These values suggest no significant difference between the observed COM PM between cluster and field.
  \item We note an offset of 28$\pm$8 \text{arcmin} between the dynamic centers ($\alpha_0$, $\delta_0$) of the cluster and field.
  \item Estimated $\Theta$ values of 122$^\circ$.22$^{\hspace{0.01cm}\scaleto{+1.48}{4.7pt}}_{\hspace{0.04cm}\scaleto{-1.51}{4.7pt}}$ for clusters and 128$^{\circ}$.02$^{\hspace{0.01cm}\scaleto{+0.79}{4.7pt}}_{\hspace{0.04cm}\scaleto{-0.81}{4.7pt}}$ for field regions point to an offset of 5$^\circ$.8$\pm$1$^\circ$.7 between them, while the inclination, $i$, remains almost similar.
  \item The modeled rotational parameters, $R_0$ and $\eta_0$ appear to be larger for field regions compared to clusters, while the $v_0$ remains almost similar.
\end{itemize}

Two major notable differences in the kinematic properties obtained are, (1) for the value of $R_0$, where clusters have a relatively low value when compared to the field population, and (2) a significant offset in the dynamic centers of cluster and field population. In the next sub-section, we look into the control population we used in our model to understand the kinematic nature of the cluster and field population.    

\subsection{Comparison with control population}\label{km_control}

Table \ref{tab3_paraemters} provides the estimated kinematic parameters for the control population as mentioned in Section \ref{sec_2}. Below we compare the cluster and field with the young MS stars and RC stars to understand their kinematic nature.

\begin{table}
    \centering

    \caption{The kinematic best-fitting parameters obtained for the LMC based on the control sample (clusters, field population, young MS stars, RC stars) are tabulated below.}
    \label{tab3_paraemters}
\renewcommand{\arraystretch}{1.5}
    \begin{tabular}{l@{\hspace{0.65em}}c@{\hspace{0.65em}}c@{\hspace{0.65em}}c@{\hspace{0.65em}}c@{\hspace{0.65em}}c@{\hspace{0.65em}}c@{\hspace{0.65em}}c@{\hspace{0.65em}}c@{\hspace{0.65em}}c@{\hspace{0.65em}}}
        \hline\hline
        \textbf{Data} & \textbf{$i$} & \textbf{$\Theta$} & \textbf{$\alpha_0$}& \textbf{$\delta_0$}& \textbf{$v_t$}& \textbf{$\theta_t$}& \textbf{$R_0$}& \textbf{$\eta$}& \textbf{$v_0$} \\
         & (deg) & (deg) & (deg) & (deg) & (km s$^{-1}$) & (deg) & (kpc) &  & (km s$^{-1}$) \\
        \hline
        Clusters& 31.11$^{\hspace{0.01cm}\scaleto{+0.82}{4.7pt}}_{\hspace{0.01cm}\scaleto{-0.81}{4.7pt}}$ & 120.27$^{\hspace{0.01cm}\scaleto{+1.65}{4.7pt}}_{\hspace{0.01cm}\scaleto{-1.64}{4.7pt}}$ & 81.71$^{\hspace{0.01cm}\scaleto{+0.13}{4.7pt}}_{\hspace{0.01cm}\scaleto{-0.13}{4.7pt}}$ & -69.6$^{\hspace{0.01cm}\scaleto{+0.03}{4.7pt}}_{\hspace{0.01cm}\scaleto{-0.03}{4.7pt}}$ & 445.01$^{\hspace{0.01cm}\scaleto{+0.58}{4.7pt}}_{\hspace{0.01cm}\scaleto{-0.59}{4.7pt}}$ & 167.86$^{\hspace{0.01cm}\scaleto{+0.22}{4.7pt}}_{\hspace{0.01cm}\scaleto{-0.22}{4.7pt}}$ & 1.63$^{\hspace{0.01cm}\scaleto{+0.11}{4.7pt}}_{\hspace{0.01cm}\scaleto{-0.11}{4.7pt}}$ & 1.88$^{\hspace{0.01cm}\scaleto{+0.17}{4.7pt}}_{\hspace{0.01cm}\scaleto{-0.15}{4.7pt}}$ & 83.95$^{\hspace{0.01cm}\scaleto{+1.3}{4.7pt}}_{\hspace{0.01cm}\scaleto{-1.3}{4.7pt}}$ \\
        \hline
        Nearby field& 30.83$^{\hspace{0.01cm}\scaleto{+0.4}{4.7pt}}_{\hspace{0.01cm}\scaleto{-0.41}{4.7pt}}$ & 127.7$^{\hspace{0.01cm}\scaleto{+0.81}{4.7pt}}_{\hspace{0.01cm}\scaleto{-0.8}{4.7pt}}$ & 80.59$^{\hspace{0.01cm}\scaleto{+0.08}{4.7pt}}_{\hspace{0.01cm}\scaleto{-0.08}{4.7pt}}$ & -69.35$^{\hspace{0.01cm}\scaleto{+0.02}{4.7pt}}_{\hspace{0.01cm}\scaleto{-0.02}{4.7pt}}$ & 442.93$^{\hspace{0.01cm}\scaleto{+0.37}{4.7pt}}_{\hspace{0.01cm}\scaleto{-0.37}{4.7pt}}$ & 169.43$^{\hspace{0.01cm}\scaleto{+0.14}{4.7pt}}_{\hspace{0.01cm}\scaleto{-0.14}{4.7pt}}$ & 2.45$^{\hspace{0.01cm}\scaleto{+0.04}{4.7pt}}_{\hspace{0.01cm}\scaleto{-0.04}{4.7pt}}$ & 2.35$^{\hspace{0.01cm}\scaleto{+0.14}{4.7pt}}_{\hspace{0.01cm}\scaleto{-0.13}{4.7pt}}$ & 84.56$^{\hspace{0.01cm}\scaleto{+1.21}{4.7pt}}_{\hspace{0.01cm}\scaleto{-1.16}{4.7pt}}$ \\
         \hline
        Young MS& 33.26$^{\hspace{0.01cm}\scaleto{+0.56}{4.7pt}}_{\hspace{0.01cm}\scaleto{-0.57}{4.7pt}}$ & 115.45$^{\hspace{0.01cm}\scaleto{+0.99}{4.7pt}}_{\hspace{0.01cm}\scaleto{-0.99}{4.7pt}}$ & 80.99$^{\hspace{0.01cm}\scaleto{+0.08}{4.7pt}}_{\hspace{0.01cm}\scaleto{-0.08}{4.7pt}}$ & -69.64$^{\hspace{0.01cm}\scaleto{+0.02}{4.7pt}}_{\hspace{0.01cm}\scaleto{-0.02}{4.7pt}}$ & 446.65$^{\hspace{0.01cm}\scaleto{+0.43}{4.7pt}}_{\hspace{0.01cm}\scaleto{-0.42}{4.7pt}}$ & 169.5$^{\hspace{0.01cm}\scaleto{+0.15}{4.7pt}}_{\hspace{0.01cm}\scaleto{-0.14}{4.7pt}}$ & 1.29$^{\hspace{0.01cm}\scaleto{+0.08}{4.7pt}}_{\hspace{0.01cm}\scaleto{-0.08}{4.7pt}}$ & 1.5$^{\hspace{0.01cm}\scaleto{+0.1}{4.7pt}}_{\hspace{0.01cm}\scaleto{-0.09}{4.7pt}}$ & 90.92$^{\hspace{0.01cm}\scaleto{+1.26}{4.7pt}}_{\hspace{0.01cm}\scaleto{-1.18}{4.7pt}}$ \\
         \hline
        Red Clump& 31.1$^{\hspace{0.01cm}\scaleto{+0.44}{4.7pt}}_{\hspace{0.01cm}\scaleto{-0.44}{4.7pt}}$ & 135.84$^{\hspace{0.01cm}\scaleto{+0.91}{4.7pt}}_{\hspace{0.01cm}\scaleto{-0.89}{4.7pt}}$ & 80.68$^{\hspace{0.01cm}\scaleto{+0.11}{4.7pt}}_{\hspace{0.01cm}\scaleto{-0.11}{4.7pt}}$ & -69.21$^{\hspace{0.01cm}\scaleto{+0.03}{4.7pt}}_{\hspace{0.01cm}\scaleto{-0.03}{4.7pt}}$ & 441.6$^{\hspace{0.01cm}\scaleto{+0.48}{4.7pt}}_{\hspace{0.01cm}\scaleto{-0.46}{4.7pt}}$ & 169.01$^{\hspace{0.01cm}\scaleto{+0.2}{4.7pt}}_{\hspace{0.01cm}\scaleto{-0.19}{4.7pt}}$ & 2.99$^{\hspace{0.01cm}\scaleto{+0.05}{4.7pt}}_{\hspace{0.01cm}\scaleto{-0.05}{4.7pt}}$ & 3.24$^{\hspace{0.01cm}\scaleto{+0.28}{4.7pt}}_{\hspace{0.01cm}\scaleto{-0.24}{4.7pt}}$ & 81.39$^{\hspace{0.01cm}\scaleto{+1.21}{4.7pt}}_{\hspace{0.01cm}\scaleto{-1.18}{4.7pt}}$ \\
        \hline\hline
    \end{tabular}
\end{table}

\begin{itemize}
    \item $\mu_{W,com}$ and $\mu_{N,com}$ appear to be similar in all the data sets of the control population without any significant difference.
    \item The value of $\alpha_0$ for the field and RC population are the same within errors, whereas the value of $\delta_0$ shows a 
 difference. The $\alpha_0$ for clusters shows a significant offset in comparison to other populations. In general, the younger population tends to show a southern and eastward dynamic center. %as noticed before (subsection \ref{clust_field_kineamtic}) between the cluster and field population. 
    \item Estimated values of $\Theta$ = 115$^{\circ}$.45$^{\hspace{0.01cm}\scaleto{+0.99}{4.7pt}}_{\hspace{0.01cm}\scaleto{-0.99}{4.7pt}}$ and $i$ = 33$^{\circ}$.26$^{\hspace{0.01cm}\scaleto{+0.56}{4.7pt}}_{\hspace{0.01cm}\scaleto{-0.57}{4.7pt}}$ for the young MS population appear distinct from other data sets in the control population. RC population has the largest value of $\Theta$ (135$^{\circ}$.84).
    \item We note a significant shift in $R_0$, $\lvert \Delta R_0 \rvert$ $\approx$ 1.7 kpc between young MS and RC population. A corresponding change of $\eta$ is observed as well. We also estimate a larger value for $v_0$ = 90.92$^{\hspace{0.01cm}\scaleto{+1.26}{4.7pt}}_{\hspace{0.01cm}\scaleto{-1.18}{4.7pt}}$ km s$^{-1}$ for the young MS population in comparison with the other data sets.
\end{itemize}
As observed in the previous comparison (subsection \ref{clust_field_kineamtic}), the value of $\Theta$ appears to be changing across the control population. Notably, the clusters and young MS population show a relatively small value in comparison to the field and RC population. Specifically, the $R_0$ for the cluster and young MS population falls within the range of 2 kpc, while the field and RC population have values exceeding 2 kpc.

\begin{figure*}
    \centering
       \includegraphics[width=1\linewidth]{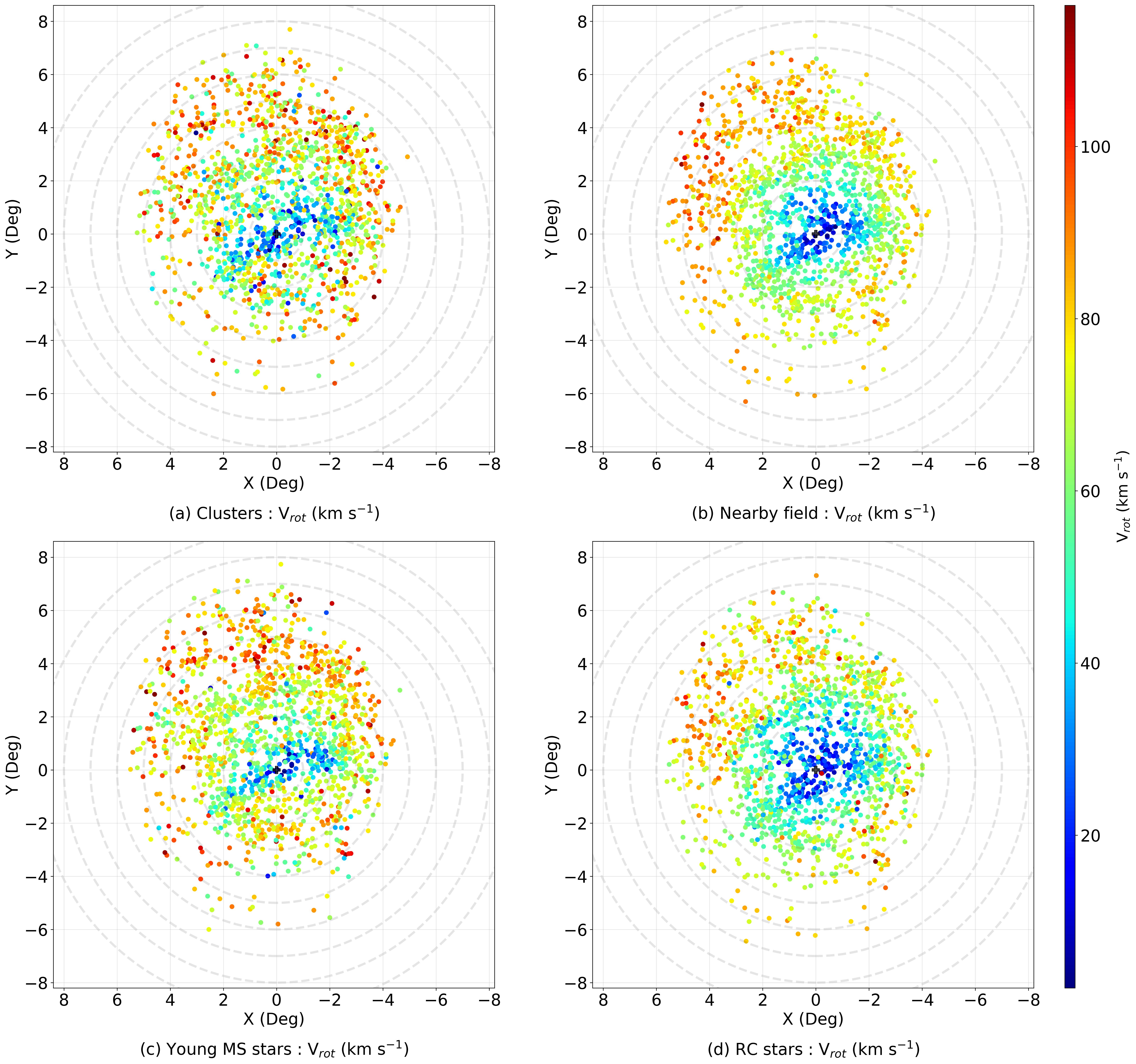}
       \caption{Spatial distribution of the rotational velocity (V$_{rot}$) of the LMC for the clusters and the three control populations are depicted here.}
       \label{fig:Rot_spatial_4_pop} 
\end{figure*}

The model parameters estimated for the cluster and the control samples are used to obtain the values of rotational velocity (V$_{rot}$) of the LMC. Figure \ref{fig:Rot_spatial_4_pop} shows the spatial distribution of V$_{rot}$ among the clusters, field stars, Young stars, and RC population. Though the overall appearance of the plots is similar, we note that clusters (panel a) and the young MS stars (panel c) appear to have similar spatial distributions of V$_{rot}$, whereas the field (panel b) and the RC stars (panel d) appear to have similar V$_{rot}$ distribution. We also note that the elongated bar feature is more prominent in the V$_{rot}$ maps of clusters and young MS stars.

This suggests that the kinematic model of the LMC depends on the age of the population. In the subsequent subsection, we look into the kinematic model of the LMC with data sets of different ages, using the cluster age groups and nearby field regions as mentioned in Section \ref{sec_2}.

\subsection{Age dependent kinematics of the LMC}\label{km_age}
The age-dependent variation of the kinematic parameters of the LMC was estimated using the data sets mentioned in Section \ref{sec_2}. Table \ref{tab4_paraemters} provides the estimated kinematic parameters for the cluster and nearby field data sets based on the cluster age groups. Below, we compare their kinematic properties.
\begin{table}
    \centering
    \caption{The kinematic best-fitting parameters obtained for the LMC based on the cluster age groups and nearby field regions are tabulated below. As mentioned in Section \ref{sec_2}, the associated field population for each cluster age group is age-wise heterogeneous.}
    \label{tab4_paraemters}
\renewcommand{\arraystretch}{1.5}
    \begin{tabular}{l@{\hspace{0.65em}}c@{\hspace{0.65em}}c@{\hspace{0.65em}}c@{\hspace{0.65em}}c@{\hspace{0.65em}}c@{\hspace{0.65em}}c@{\hspace{0.65em}}c@{\hspace{0.65em}}c@{\hspace{0.65em}}c@{\hspace{0.65em}}}
        \hline\hline
        \textbf{Data} & \textbf{$i$} & \textbf{$\Theta$} & \textbf{$\alpha_0$}& \textbf{$\delta_0$}& \textbf{$v_t$}& \textbf{$\theta_t$}& \textbf{$R_0$}& \textbf{$\eta$}& \textbf{$v_0$} \\
         & (deg) & (deg) & (deg) & (deg) & (km s$^{-1}$) & (deg) & (kpc) &  & (km s$^{-1}$) \\
        \hline
        C$_{AG-1}$& 31.56$^{\hspace{0.01cm}\scaleto{+1.22}{4.7pt}}_{\hspace{0.01cm}\scaleto{-1.3}{4.7pt}}$ & 145.25$^{\hspace{0.01cm}\scaleto{+2.93}{4.7pt}}_{\hspace{0.01cm}\scaleto{-3.01}{4.7pt}}$ & 80.84$^{\hspace{0.01cm}\scaleto{+0.38}{4.7pt}}_{\hspace{0.01cm}\scaleto{-0.38}{4.7pt}}$ & -69.19$^{\hspace{0.01cm}\scaleto{+0.13}{4.7pt}}_{\hspace{0.01cm}\scaleto{-0.14}{4.7pt}}$ & 440.57$^{\hspace{0.01cm}\scaleto{+1.49}{4.7pt}}_{\hspace{0.01cm}\scaleto{-1.52}{4.7pt}}$ & 169.02$^{\hspace{0.01cm}\scaleto{+0.63}{4.7pt}}_{\hspace{0.01cm}\scaleto{-0.63}{4.7pt}}$ & 3.14$^{\hspace{0.01cm}\scaleto{+0.18}{4.7pt}}_{\hspace{0.01cm}\scaleto{-0.22}{4.7pt}}$ & 5.61$^{\hspace{0.01cm}\scaleto{+1.76}{4.7pt}}_{\hspace{0.01cm}\scaleto{-1.4}{4.7pt}}$ & 78.63$^{\hspace{0.01cm}\scaleto{+1.4}{4.7pt}}_{\hspace{0.01cm}\scaleto{-1.33}{4.7pt}}$ \\
        F$_{AG-1}$& 31.11$^{\hspace{0.01cm}\scaleto{+0.78}{4.7pt}}_{\hspace{0.01cm}\scaleto{-0.79}{4.7pt}}$ & 137.44$^{\hspace{0.01cm}\scaleto{+1.67}{4.7pt}}_{\hspace{0.01cm}\scaleto{-1.67}{4.7pt}}$ & 80.68$^{\hspace{0.01cm}\scaleto{+0.18}{4.7pt}}_{\hspace{0.01cm}\scaleto{-0.19}{4.7pt}}$ & -69.19$^{\hspace{0.01cm}\scaleto{+0.06}{4.7pt}}_{\hspace{0.01cm}\scaleto{-0.06}{4.7pt}}$ & 439.83$^{\hspace{0.01cm}\scaleto{+0.76}{4.7pt}}_{\hspace{0.01cm}\scaleto{-0.75}{4.7pt}}$ & 169.14$^{\hspace{0.01cm}\scaleto{+0.33}{4.7pt}}_{\hspace{0.01cm}\scaleto{-0.32}{4.7pt}}$ & 2.84$^{\hspace{0.01cm}\scaleto{+0.08}{4.7pt}}_{\hspace{0.01cm}\scaleto{-0.08}{4.7pt}}$ & 5.44$^{\hspace{0.01cm}\scaleto{+0.97}{4.7pt}}_{\hspace{0.01cm}\scaleto{-0.81}{4.7pt}}$ & 77.21$^{\hspace{0.01cm}\scaleto{+1.12}{4.7pt}}_{\hspace{0.01cm}\scaleto{-1.04}{4.7pt}}$ \\
        \hline
        C$_{AG-2}$& 28.33$^{\hspace{0.01cm}\scaleto{+1.33}{4.7pt}}_{\hspace{0.01cm}\scaleto{-1.45}{4.7pt}}$ & 133.87$^{\hspace{0.01cm}\scaleto{+3.34}{4.7pt}}_{\hspace{0.01cm}\scaleto{-3.42}{4.7pt}}$ & 81.64$^{\hspace{0.01cm}\scaleto{+0.27}{4.7pt}}_{\hspace{0.01cm}\scaleto{-0.27}{4.7pt}}$ & -69.6$^{\hspace{0.01cm}\scaleto{+0.07}{4.7pt}}_{\hspace{0.01cm}\scaleto{-0.07}{4.7pt}}$ & 445.16$^{\hspace{0.01cm}\scaleto{+1.17}{4.7pt}}_{\hspace{0.01cm}\scaleto{-1.11}{4.7pt}}$ & 168.04$^{\hspace{0.01cm}\scaleto{+0.46}{4.7pt}}_{\hspace{0.01cm}\scaleto{-0.45}{4.7pt}}$ & 2.17$^{\hspace{0.01cm}\scaleto{+0.2}{4.7pt}}_{\hspace{0.01cm}\scaleto{-0.21}{4.7pt}}$ & 2.52$^{\hspace{0.01cm}\scaleto{+0.46}{4.7pt}}_{\hspace{0.01cm}\scaleto{-0.36}{4.7pt}}$ & 80.63$^{\hspace{0.01cm}\scaleto{+1.52}{4.7pt}}_{\hspace{0.01cm}\scaleto{-1.51}{4.7pt}}$ \\
        F$_{AG-2}$& 30.79$^{\hspace{0.01cm}\scaleto{+0.72}{4.7pt}}_{\hspace{0.01cm}\scaleto{-0.72}{4.7pt}}$ & 130.17$^{\hspace{0.01cm}\scaleto{+1.61}{4.7pt}}_{\hspace{0.01cm}\scaleto{-1.63}{4.7pt}}$ & 80.7$^{\hspace{0.01cm}\scaleto{+0.15}{4.7pt}}_{\hspace{0.01cm}\scaleto{-0.15}{4.7pt}}$ & -69.25$^{\hspace{0.01cm}\scaleto{+0.04}{4.7pt}}_{\hspace{0.01cm}\scaleto{-0.04}{4.7pt}}$ & 442.59$^{\hspace{0.01cm}\scaleto{+0.62}{4.7pt}}_{\hspace{0.01cm}\scaleto{-0.63}{4.7pt}}$ & 169.18$^{\hspace{0.01cm}\scaleto{+0.27}{4.7pt}}_{\hspace{0.01cm}\scaleto{-0.27}{4.7pt}}$ & 2.52$^{\hspace{0.01cm}\scaleto{+0.07}{4.7pt}}_{\hspace{0.01cm}\scaleto{-0.08}{4.7pt}}$ & 2.87$^{\hspace{0.01cm}\scaleto{+0.29}{4.7pt}}_{\hspace{0.01cm}\scaleto{-0.25}{4.7pt}}$ & 81.03$^{\hspace{0.01cm}\scaleto{+1.3}{4.7pt}}_{\hspace{0.01cm}\scaleto{-1.28}{4.7pt}}$ \\
        \hline
        C$_{AG-3}$& 35.36$^{\hspace{0.01cm}\scaleto{+1.2}{4.7pt}}_{\hspace{0.01cm}\scaleto{-1.31}{4.7pt}}$ & 111.11$^{\hspace{0.01cm}\scaleto{+2.08}{4.7pt}}_{\hspace{0.01cm}\scaleto{-2.04}{4.7pt}}$ & 81.59$^{\hspace{0.01cm}\scaleto{+0.21}{4.7pt}}_{\hspace{0.01cm}\scaleto{-0.21}{4.7pt}}$ & -69.6$^{\hspace{0.01cm}\scaleto{+0.05}{4.7pt}}_{\hspace{0.01cm}\scaleto{-0.05}{4.7pt}}$ & 445.2$^{\hspace{0.01cm}\scaleto{+0.9}{4.7pt}}_{\hspace{0.01cm}\scaleto{-0.9}{4.7pt}}$ & 168.31$^{\hspace{0.01cm}\scaleto{+0.36}{4.7pt}}_{\hspace{0.01cm}\scaleto{-0.35}{4.7pt}}$ & 1.34$^{\hspace{0.01cm}\scaleto{+0.17}{4.7pt}}_{\hspace{0.01cm}\scaleto{-0.16}{4.7pt}}$ & 2.06$^{\hspace{0.01cm}\scaleto{+0.34}{4.7pt}}_{\hspace{0.01cm}\scaleto{-0.27}{4.7pt}}$ & 81.31$^{\hspace{0.01cm}\scaleto{+1.51}{4.7pt}}_{\hspace{0.01cm}\scaleto{-1.49}{4.7pt}}$ \\
        F$_{AG-3}$& 29.89$^{\hspace{0.01cm}\scaleto{+0.72}{4.7pt}}_{\hspace{0.01cm}\scaleto{-0.73}{4.7pt}}$ & 123.5$^{\hspace{0.01cm}\scaleto{+1.41}{4.7pt}}_{\hspace{0.01cm}\scaleto{-1.46}{4.7pt}}$ & 80.36$^{\hspace{0.01cm}\scaleto{+0.12}{4.7pt}}_{\hspace{0.01cm}\scaleto{-0.12}{4.7pt}}$ & -69.34$^{\hspace{0.01cm}\scaleto{+0.03}{4.7pt}}_{\hspace{0.01cm}\scaleto{-0.03}{4.7pt}}$ & 443.01$^{\hspace{0.01cm}\scaleto{+0.56}{4.7pt}}_{\hspace{0.01cm}\scaleto{-0.55}{4.7pt}}$ & 169.84$^{\hspace{0.01cm}\scaleto{+0.22}{4.7pt}}_{\hspace{0.01cm}\scaleto{-0.22}{4.7pt}}$ & 2.28$^{\hspace{0.01cm}\scaleto{+0.06}{4.7pt}}_{\hspace{0.01cm}\scaleto{-0.07}{4.7pt}}$ & 2.34$^{\hspace{0.01cm}\scaleto{+0.18}{4.7pt}}_{\hspace{0.01cm}\scaleto{-0.16}{4.7pt}}$ & 81.88$^{\hspace{0.01cm}\scaleto{+1.47}{4.7pt}}_{\hspace{0.01cm}\scaleto{-1.43}{4.7pt}}$ \\
        \hline
        C$_{AG-4}$& 33.34$^{\hspace{0.01cm}\scaleto{+1.87}{4.7pt}}_{\hspace{0.01cm}\scaleto{-2.04}{4.7pt}}$ & 118.39$^{\hspace{0.01cm}\scaleto{+2.98}{4.7pt}}_{\hspace{0.01cm}\scaleto{-3.16}{4.7pt}}$ & 81.29$^{\hspace{0.01cm}\scaleto{+0.26}{4.7pt}}_{\hspace{0.01cm}\scaleto{-0.26}{4.7pt}}$ & -69.68$^{\hspace{0.01cm}\scaleto{+0.08}{4.7pt}}_{\hspace{0.01cm}\scaleto{-0.08}{4.7pt}}$ & 444.4$^{\hspace{0.01cm}\scaleto{+1.65}{4.7pt}}_{\hspace{0.01cm}\scaleto{-1.59}{4.7pt}}$ & 168.33$^{\hspace{0.01cm}\scaleto{+0.46}{4.7pt}}_{\hspace{0.01cm}\scaleto{-0.46}{4.7pt}}$ & 1.98$^{\hspace{0.01cm}\scaleto{+0.17}{4.7pt}}_{\hspace{0.01cm}\scaleto{-0.18}{4.7pt}}$ & 3.69$^{\hspace{0.01cm}\scaleto{+0.85}{4.7pt}}_{\hspace{0.01cm}\scaleto{-0.64}{4.7pt}}$ & 79.6$^{\hspace{0.01cm}\scaleto{+1.62}{4.7pt}}_{\hspace{0.01cm}\scaleto{-1.56}{4.7pt}}$ \\
        F$_{AG-4}$& 30.38$^{\hspace{0.01cm}\scaleto{+1.06}{4.7pt}}_{\hspace{0.01cm}\scaleto{-1.1}{4.7pt}}$ & 125.54$^{\hspace{0.01cm}\scaleto{+1.75}{4.7pt}}_{\hspace{0.01cm}\scaleto{-1.84}{4.7pt}}$ & 80.6$^{\hspace{0.01cm}\scaleto{+0.2}{4.7pt}}_{\hspace{0.01cm}\scaleto{-0.2}{4.7pt}}$ & -69.57$^{\hspace{0.01cm}\scaleto{+0.06}{4.7pt}}_{\hspace{0.01cm}\scaleto{-0.06}{4.7pt}}$ & 445.6$^{\hspace{0.01cm}\scaleto{+1.14}{4.7pt}}_{\hspace{0.01cm}\scaleto{-1.13}{4.7pt}}$ & 169.5$^{\hspace{0.01cm}\scaleto{+0.37}{4.7pt}}_{\hspace{0.01cm}\scaleto{-0.36}{4.7pt}}$ & 2.54$^{\hspace{0.01cm}\scaleto{+0.1}{4.7pt}}_{\hspace{0.01cm}\scaleto{-0.11}{4.7pt}}$ & 3.1$^{\hspace{0.01cm}\scaleto{+0.4}{4.7pt}}_{\hspace{0.01cm}\scaleto{-0.35}{4.7pt}}$ & 81.96$^{\hspace{0.01cm}\scaleto{+1.54}{4.7pt}}_{\hspace{0.01cm}\scaleto{-1.5}{4.7pt}}$ \\
        \hline
        \hline
    \end{tabular}
\end{table}
\begin{itemize}
    \item As noted in the previous subsections, the COM PM shows minimal variations across different age groups. Even slight variations in the parameter $v_t$ estimated as in Table \ref{tab4_paraemters} for different age groups do not result in substantial shifts in $\mu_{W,com}$ and $\mu_{N,com}$.
    \item The clusters ranging in age from $\sim$ 100 Myr to 1.25 Gyr show a small offset in the ($\alpha_0$, $\delta_0$) to the South-East with respect to the older clusters, as evident from C$_{AG-2}$ and C$_{AG-3}$. Meanwhile, the nearby field population in these age groups shows variation only in the South. 
    \item $\Theta$ attains its minimum value of 111$^{\circ}$.11$^{\hspace{0.01cm}\scaleto{+2.08}{4.7pt}}_{\hspace{0.01cm}\scaleto{-2.04}{4.7pt}}$ for the C$_{AG-3}$, with field also showing a similar trend. Notably, the $i$ (35$^{\circ}$.36$^{\hspace{0.01cm}\scaleto{+1.2}{4.7pt}}_{\hspace{0.01cm}\scaleto{-1.31}{4.7pt}}$) for C$_{AG-3}$ is the largest when compared to other age groups. In the case of the field regions, we do not detect any significant shift in $i$, but a smaller shift in $\Theta$ is noted.
    \item In the cluster data sets, the minimum value of $R_0$ is estimated to be 1.34$^{\hspace{0.01cm}\scaleto{+0.17}{4.7pt}}_{\hspace{0.01cm}\scaleto{-0.16}{4.7pt}}$ kpc for the C$_{AG-3}$, while its maximum value occurs at 3.14$^{\hspace{0.01cm}\scaleto{+0.18}{4.7pt}}_{\hspace{0.01cm}\scaleto{-0.22}{4.7pt}}$ kpc for the C$_{AG-1}$. Similarly, in the field data sets, there is a comparable pattern with the minimum and maximum values of 2.28$^{\hspace{0.01cm}\scaleto{+0.06}{4.7pt}}_{\hspace{0.01cm}\scaleto{-0.07}{4.7pt}}$ kpc and 2.84$^{\hspace{0.01cm}\scaleto{+0.08}{4.7pt}}_{\hspace{0.01cm}\scaleto{-0.08}{4.7pt}}$ kpc for F$_{AG-3}$ and F$_{AG-1}$, respectively. $\eta$ also shows a similar trend in the age-based data sets. 
    \item The values of $v_0$ remain relatively consistent across all datasets, and any observed changes are not significant, as they are within the margin of errors.
\end{itemize}
 
 As noted in the above comparison, there is a noticeable deviation of kinematic parameters for the cluster age groups, in contrast to the datasets of the field region associated with each age group. The comparison presented in this section has shown that the dependence of kinematic parameters on the age of the population is indeed present, and the variations are statistically significant. In the following subsections, we further investigate the dependence of estimated parameters on the cluster richness as mentioned in Section \ref{sec_2}.

\subsection{Influence of cluster richness on the kinematic model}
Table \ref{tab5_paraemters} shows the estimated kinematic parameters for the data sets based on cluster richness, as outlined in Section \ref{sec_2}. The cluster data sets and the nearby field regions are modeled based on cluster richness, which helped in checking for any kinematic changes in the model based on poor and rich clusters.
\begin{table}
    \centering
    \caption{The kinematic best-fitting parameters obtained for the LMC based on the cluster richness and their nearby field regions are tabulated below.}
    \label{tab5_paraemters}
\renewcommand{\arraystretch}{1.5}
    \begin{tabular}{l@{\hspace{0.65em}}c@{\hspace{0.65em}}c@{\hspace{0.65em}}c@{\hspace{0.65em}}c@{\hspace{0.65em}}c@{\hspace{0.65em}}c@{\hspace{0.65em}}c@{\hspace{0.65em}}c@{\hspace{0.65em}}c@{\hspace{0.65em}}}
        \hline\hline
        \textbf{Data} & \textbf{$i$} & \textbf{$\Theta$} & \textbf{$\alpha_0$}& \textbf{$\delta_0$}& \textbf{$v_t$}& \textbf{$\theta_t$}& \textbf{$R_0$}& \textbf{$\eta$}& \textbf{$v_0$} \\
         & (deg) & (deg) & (deg) & (deg) & (km s$^{-1}$) & (deg) & (kpc) &  & (km s$^{-1}$) \\
        \hline
        C$_{10}$& 31.78$^{\hspace{0.01cm}\scaleto{+0.72}{4.7pt}}_{\hspace{0.01cm}\scaleto{-0.75}{4.7pt}}$ & 122.19$^{\hspace{0.01cm}\scaleto{+1.48}{4.7pt}}_{\hspace{0.01cm}\scaleto{-1.52}{4.7pt}}$ & 81.7$^{\hspace{0.01cm}\scaleto{+0.13}{4.7pt}}_{\hspace{0.01cm}\scaleto{-0.13}{4.7pt}}$ & -69.61$^{\hspace{0.01cm}\scaleto{+0.03}{4.7pt}}_{\hspace{0.01cm}\scaleto{-0.03}{4.7pt}}$ & 444.87$^{\hspace{0.01cm}\scaleto{+0.55}{4.7pt}}_{\hspace{0.01cm}\scaleto{-0.54}{4.7pt}}$ & 167.87$^{\hspace{0.01cm}\scaleto{+0.22}{4.7pt}}_{\hspace{0.01cm}\scaleto{-0.22}{4.7pt}}$ & 1.59$^{\hspace{0.01cm}\scaleto{+0.11}{4.7pt}}_{\hspace{0.01cm}\scaleto{-0.11}{4.7pt}}$ & 1.85$^{\hspace{0.01cm}\scaleto{+0.17}{4.7pt}}_{\hspace{0.01cm}\scaleto{-0.15}{4.7pt}}$ & 83.99$^{\hspace{0.01cm}\scaleto{+1.3}{4.7pt}}_{\hspace{0.01cm}\scaleto{-1.24}{4.7pt}}$ \\
        F$_{10}$& 30.78$^{\hspace{0.01cm}\scaleto{+0.4}{4.7pt}}_{\hspace{0.01cm}\scaleto{-0.41}{4.7pt}}$ & 127.85$^{\hspace{0.01cm}\scaleto{+0.82}{4.7pt}}_{\hspace{0.01cm}\scaleto{-0.79}{4.7pt}}$ & 80.59$^{\hspace{0.01cm}\scaleto{+0.07}{4.7pt}}_{\hspace{0.01cm}\scaleto{-0.08}{4.7pt}}$ & -69.35$^{\hspace{0.01cm}\scaleto{+0.02}{4.7pt}}_{\hspace{0.01cm}\scaleto{-0.02}{4.7pt}}$ & 442.82$^{\hspace{0.01cm}\scaleto{+0.35}{4.7pt}}_{\hspace{0.01cm}\scaleto{-0.35}{4.7pt}}$ & 169.43$^{\hspace{0.01cm}\scaleto{+0.14}{4.7pt}}_{\hspace{0.01cm}\scaleto{-0.14}{4.7pt}}$ & 2.41$^{\hspace{0.01cm}\scaleto{+0.04}{4.7pt}}_{\hspace{0.01cm}\scaleto{-0.04}{4.7pt}}$ & 2.34$^{\hspace{0.01cm}\scaleto{+0.14}{4.7pt}}_{\hspace{0.01cm}\scaleto{-0.13}{4.7pt}}$ & 84.15$^{\hspace{0.01cm}\scaleto{+1.17}{4.7pt}}_{\hspace{0.01cm}\scaleto{-1.14}{4.7pt}}$ \\
        \hline
        C$_{20}$& 31.76$^{\hspace{0.01cm}\scaleto{+0.85}{4.7pt}}_{\hspace{0.01cm}\scaleto{-0.88}{4.7pt}}$ & 119.48$^{\hspace{0.01cm}\scaleto{+1.73}{4.7pt}}_{\hspace{0.01cm}\scaleto{-1.73}{4.7pt}}$ & 81.85$^{\hspace{0.01cm}\scaleto{+0.14}{4.7pt}}_{\hspace{0.01cm}\scaleto{-0.14}{4.7pt}}$ & -69.62$^{\hspace{0.01cm}\scaleto{+0.03}{4.7pt}}_{\hspace{0.01cm}\scaleto{-0.03}{4.7pt}}$ & 444.81$^{\hspace{0.01cm}\scaleto{+0.58}{4.7pt}}_{\hspace{0.01cm}\scaleto{-0.61}{4.7pt}}$ & 167.59$^{\hspace{0.01cm}\scaleto{+0.24}{4.7pt}}_{\hspace{0.01cm}\scaleto{-0.23}{4.7pt}}$ & 1.47$^{\hspace{0.01cm}\scaleto{+0.12}{4.7pt}}_{\hspace{0.01cm}\scaleto{-0.12}{4.7pt}}$ & 1.81$^{\hspace{0.01cm}\scaleto{+0.18}{4.7pt}}_{\hspace{0.01cm}\scaleto{-0.16}{4.7pt}}$ & 82.89$^{\hspace{0.01cm}\scaleto{+1.34}{4.7pt}}_{\hspace{0.01cm}\scaleto{-1.33}{4.7pt}}$ \\
        F$_{20}$& 30.18$^{\hspace{0.01cm}\scaleto{+0.49}{4.7pt}}_{\hspace{0.01cm}\scaleto{-0.49}{4.7pt}}$ & 127.05$^{\hspace{0.01cm}\scaleto{+0.94}{4.7pt}}_{\hspace{0.01cm}\scaleto{-0.97}{4.7pt}}$ & 80.59$^{\hspace{0.01cm}\scaleto{+0.09}{4.7pt}}_{\hspace{0.01cm}\scaleto{-0.09}{4.7pt}}$ & -69.34$^{\hspace{0.01cm}\scaleto{+0.02}{4.7pt}}_{\hspace{0.01cm}\scaleto{-0.02}{4.7pt}}$ & 442.55$^{\hspace{0.01cm}\scaleto{+0.4}{4.7pt}}_{\hspace{0.01cm}\scaleto{-0.4}{4.7pt}}$ & 169.43$^{\hspace{0.01cm}\scaleto{+0.16}{4.7pt}}_{\hspace{0.01cm}\scaleto{-0.16}{4.7pt}}$ & 2.41$^{\hspace{0.01cm}\scaleto{+0.05}{4.7pt}}_{\hspace{0.01cm}\scaleto{-0.05}{4.7pt}}$ & 2.55$^{\hspace{0.01cm}\scaleto{+0.18}{4.7pt}}_{\hspace{0.01cm}\scaleto{-0.16}{4.7pt}}$ & 82.22$^{\hspace{0.01cm}\scaleto{+1.19}{4.7pt}}_{\hspace{0.01cm}\scaleto{-1.15}{4.7pt}}$ \\
        \hline
        C$_{30}$& 31.62$^{\hspace{0.01cm}\scaleto{+0.94}{4.7pt}}_{\hspace{0.01cm}\scaleto{-0.98}{4.7pt}}$ & 119.72$^{\hspace{0.01cm}\scaleto{+1.91}{4.7pt}}_{\hspace{0.01cm}\scaleto{-1.94}{4.7pt}}$ & 82.06$^{\hspace{0.01cm}\scaleto{+0.14}{4.7pt}}_{\hspace{0.01cm}\scaleto{-0.14}{4.7pt}}$ & -69.63$^{\hspace{0.01cm}\scaleto{+0.04}{4.7pt}}_{\hspace{0.01cm}\scaleto{-0.04}{4.7pt}}$ & 445.06$^{\hspace{0.01cm}\scaleto{+0.65}{4.7pt}}_{\hspace{0.01cm}\scaleto{-0.64}{4.7pt}}$ & 167.23$^{\hspace{0.01cm}\scaleto{+0.24}{4.7pt}}_{\hspace{0.01cm}\scaleto{-0.24}{4.7pt}}$ & 1.31$^{\hspace{0.01cm}\scaleto{+0.14}{4.7pt}}_{\hspace{0.01cm}\scaleto{-0.13}{4.7pt}}$ & 1.69$^{\hspace{0.01cm}\scaleto{+0.18}{4.7pt}}_{\hspace{0.01cm}\scaleto{-0.16}{4.7pt}}$ & 82.59$^{\hspace{0.01cm}\scaleto{+1.39}{4.7pt}}_{\hspace{0.01cm}\scaleto{-1.39}{4.7pt}}$ \\
        F$_{30}$& 29.86$^{\hspace{0.01cm}\scaleto{+0.55}{4.7pt}}_{\hspace{0.01cm}\scaleto{-0.55}{4.7pt}}$ & 127.06$^{\hspace{0.01cm}\scaleto{+1.13}{4.7pt}}_{\hspace{0.01cm}\scaleto{-1.13}{4.7pt}}$ & 80.72$^{\hspace{0.01cm}\scaleto{+0.1}{4.7pt}}_{\hspace{0.01cm}\scaleto{-0.1}{4.7pt}}$ & -69.33$^{\hspace{0.01cm}\scaleto{+0.03}{4.7pt}}_{\hspace{0.01cm}\scaleto{-0.03}{4.7pt}}$ & 442.4$^{\hspace{0.01cm}\scaleto{+0.46}{4.7pt}}_{\hspace{0.01cm}\scaleto{-0.46}{4.7pt}}$ & 169.19$^{\hspace{0.01cm}\scaleto{+0.19}{4.7pt}}_{\hspace{0.01cm}\scaleto{-0.18}{4.7pt}}$ & 2.42$^{\hspace{0.01cm}\scaleto{+0.05}{4.7pt}}_{\hspace{0.01cm}\scaleto{-0.05}{4.7pt}}$ & 2.58$^{\hspace{0.01cm}\scaleto{+0.19}{4.7pt}}_{\hspace{0.01cm}\scaleto{-0.18}{4.7pt}}$ & 81.9$^{\hspace{0.01cm}\scaleto{+1.27}{4.7pt}}_{\hspace{0.01cm}\scaleto{-1.24}{4.7pt}}$ \\
        \hline
        C$_{40}$& 31.04$^{\hspace{0.01cm}\scaleto{+1.04}{4.7pt}}_{\hspace{0.01cm}\scaleto{-1.08}{4.7pt}}$ & 118.5$^{\hspace{0.01cm}\scaleto{+2.09}{4.7pt}}_{\hspace{0.01cm}\scaleto{-2.12}{4.7pt}}$ & 82.2$^{\hspace{0.01cm}\scaleto{+0.16}{4.7pt}}_{\hspace{0.01cm}\scaleto{-0.17}{4.7pt}}$ & -69.61$^{\hspace{0.01cm}\scaleto{+0.04}{4.7pt}}_{\hspace{0.01cm}\scaleto{-0.04}{4.7pt}}$ & 445.07$^{\hspace{0.01cm}\scaleto{+0.7}{4.7pt}}_{\hspace{0.01cm}\scaleto{-0.66}{4.7pt}}$ & 166.97$^{\hspace{0.01cm}\scaleto{+0.28}{4.7pt}}_{\hspace{0.01cm}\scaleto{-0.28}{4.7pt}}$ & 1.34$^{\hspace{0.01cm}\scaleto{+0.14}{4.7pt}}_{\hspace{0.01cm}\scaleto{-0.15}{4.7pt}}$ & 1.68$^{\hspace{0.01cm}\scaleto{+0.19}{4.7pt}}_{\hspace{0.01cm}\scaleto{-0.17}{4.7pt}}$ & 82.04$^{\hspace{0.01cm}\scaleto{+1.47}{4.7pt}}_{\hspace{0.01cm}\scaleto{-1.42}{4.7pt}}$ \\
        F$_{40}$& 28.81$^{\hspace{0.01cm}\scaleto{+0.63}{4.7pt}}_{\hspace{0.01cm}\scaleto{-0.64}{4.7pt}}$ & 126.27$^{\hspace{0.01cm}\scaleto{+1.31}{4.7pt}}_{\hspace{0.01cm}\scaleto{-1.37}{4.7pt}}$ & 80.82$^{\hspace{0.01cm}\scaleto{+0.11}{4.7pt}}_{\hspace{0.01cm}\scaleto{-0.11}{4.7pt}}$ & -69.33$^{\hspace{0.01cm}\scaleto{+0.03}{4.7pt}}_{\hspace{0.01cm}\scaleto{-0.03}{4.7pt}}$ & 442.28$^{\hspace{0.01cm}\scaleto{+0.5}{4.7pt}}_{\hspace{0.01cm}\scaleto{-0.51}{4.7pt}}$ & 169.04$^{\hspace{0.01cm}\scaleto{+0.2}{4.7pt}}_{\hspace{0.01cm}\scaleto{-0.2}{4.7pt}}$ & 2.35$^{\hspace{0.01cm}\scaleto{+0.06}{4.7pt}}_{\hspace{0.01cm}\scaleto{-0.06}{4.7pt}}$ & 2.52$^{\hspace{0.01cm}\scaleto{+0.2}{4.7pt}}_{\hspace{0.01cm}\scaleto{-0.19}{4.7pt}}$ & 80.96$^{\hspace{0.01cm}\scaleto{+1.32}{4.7pt}}_{\hspace{0.01cm}\scaleto{-1.26}{4.7pt}}$ \\
        \hline
        C$_{50}$& 31.05$^{\hspace{0.01cm}\scaleto{+1.13}{4.7pt}}_{\hspace{0.01cm}\scaleto{-1.24}{4.7pt}}$ & 116.34$^{\hspace{0.01cm}\scaleto{+2.27}{4.7pt}}_{\hspace{0.01cm}\scaleto{-2.3}{4.7pt}}$ & 82.36$^{\hspace{0.01cm}\scaleto{+0.17}{4.7pt}}_{\hspace{0.01cm}\scaleto{-0.17}{4.7pt}}$ & -69.65$^{\hspace{0.01cm}\scaleto{+0.04}{4.7pt}}_{\hspace{0.01cm}\scaleto{-0.04}{4.7pt}}$ & 445.58$^{\hspace{0.01cm}\scaleto{+0.73}{4.7pt}}_{\hspace{0.01cm}\scaleto{-0.74}{4.7pt}}$ & 166.76$^{\hspace{0.01cm}\scaleto{+0.29}{4.7pt}}_{\hspace{0.01cm}\scaleto{-0.29}{4.7pt}}$ & 1.15$^{\hspace{0.01cm}\scaleto{+0.15}{4.7pt}}_{\hspace{0.01cm}\scaleto{-0.15}{4.7pt}}$ & 1.59$^{\hspace{0.01cm}\scaleto{+0.19}{4.7pt}}_{\hspace{0.01cm}\scaleto{-0.17}{4.7pt}}$ & 81.29$^{\hspace{0.01cm}\scaleto{+1.52}{4.7pt}}_{\hspace{0.01cm}\scaleto{-1.47}{4.7pt}}$ \\
        F$_{50}$& 28.94$^{\hspace{0.01cm}\scaleto{+0.69}{4.7pt}}_{\hspace{0.01cm}\scaleto{-0.72}{4.7pt}}$ & 126.19$^{\hspace{0.01cm}\scaleto{+1.5}{4.7pt}}_{\hspace{0.01cm}\scaleto{-1.53}{4.7pt}}$ & 81.06$^{\hspace{0.01cm}\scaleto{+0.12}{4.7pt}}_{\hspace{0.01cm}\scaleto{-0.12}{4.7pt}}$ & -69.37$^{\hspace{0.01cm}\scaleto{+0.03}{4.7pt}}_{\hspace{0.01cm}\scaleto{-0.03}{4.7pt}}$ & 442.56$^{\hspace{0.01cm}\scaleto{+0.51}{4.7pt}}_{\hspace{0.01cm}\scaleto{-0.53}{4.7pt}}$ & 168.66$^{\hspace{0.01cm}\scaleto{+0.21}{4.7pt}}_{\hspace{0.01cm}\scaleto{-0.21}{4.7pt}}$ & 2.18$^{\hspace{0.01cm}\scaleto{+0.07}{4.7pt}}_{\hspace{0.01cm}\scaleto{-0.07}{4.7pt}}$ & 2.37$^{\hspace{0.01cm}\scaleto{+0.21}{4.7pt}}_{\hspace{0.01cm}\scaleto{-0.18}{4.7pt}}$ & 80.25$^{\hspace{0.01cm}\scaleto{+1.38}{4.7pt}}_{\hspace{0.01cm}\scaleto{-1.31}{4.7pt}}$ \\
        \hline
        C$_{60}$& 29.75$^{\hspace{0.01cm}\scaleto{+1.35}{4.7pt}}_{\hspace{0.01cm}\scaleto{-1.4}{4.7pt}}$ & 116.94$^{\hspace{0.01cm}\scaleto{+2.6}{4.7pt}}_{\hspace{0.01cm}\scaleto{-2.69}{4.7pt}}$ & 82.5$^{\hspace{0.01cm}\scaleto{+0.18}{4.7pt}}_{\hspace{0.01cm}\scaleto{-0.17}{4.7pt}}$ & -69.69$^{\hspace{0.01cm}\scaleto{+0.05}{4.7pt}}_{\hspace{0.01cm}\scaleto{-0.04}{4.7pt}}$ & 445.95$^{\hspace{0.01cm}\scaleto{+0.77}{4.7pt}}_{\hspace{0.01cm}\scaleto{-0.75}{4.7pt}}$ & 166.49$^{\hspace{0.01cm}\scaleto{+0.29}{4.7pt}}_{\hspace{0.01cm}\scaleto{-0.3}{4.7pt}}$ & 0.95$^{\hspace{0.01cm}\scaleto{+0.16}{4.7pt}}_{\hspace{0.01cm}\scaleto{-0.15}{4.7pt}}$ & 1.43$^{\hspace{0.01cm}\scaleto{+0.19}{4.7pt}}_{\hspace{0.01cm}\scaleto{-0.16}{4.7pt}}$ & 80.06$^{\hspace{0.01cm}\scaleto{+1.56}{4.7pt}}_{\hspace{0.01cm}\scaleto{-1.58}{4.7pt}}$ \\
        F$_{60}$& 28.13$^{\hspace{0.01cm}\scaleto{+0.77}{4.7pt}}_{\hspace{0.01cm}\scaleto{-0.8}{4.7pt}}$ & 127.28$^{\hspace{0.01cm}\scaleto{+1.76}{4.7pt}}_{\hspace{0.01cm}\scaleto{-1.77}{4.7pt}}$ & 81.34$^{\hspace{0.01cm}\scaleto{+0.12}{4.7pt}}_{\hspace{0.01cm}\scaleto{-0.12}{4.7pt}}$ & -69.42$^{\hspace{0.01cm}\scaleto{+0.04}{4.7pt}}_{\hspace{0.01cm}\scaleto{-0.04}{4.7pt}}$ & 442.91$^{\hspace{0.01cm}\scaleto{+0.58}{4.7pt}}_{\hspace{0.01cm}\scaleto{-0.56}{4.7pt}}$ & 168.16$^{\hspace{0.01cm}\scaleto{+0.22}{4.7pt}}_{\hspace{0.01cm}\scaleto{-0.22}{4.7pt}}$ & 2.07$^{\hspace{0.01cm}\scaleto{+0.08}{4.7pt}}_{\hspace{0.01cm}\scaleto{-0.08}{4.7pt}}$ & 2.12$^{\hspace{0.01cm}\scaleto{+0.19}{4.7pt}}_{\hspace{0.01cm}\scaleto{-0.17}{4.7pt}}$ & 80.47$^{\hspace{0.01cm}\scaleto{+1.44}{4.7pt}}_{\hspace{0.01cm}\scaleto{-1.46}{4.7pt}}$ \\
        \hline
        \hline
    \end{tabular}
\end{table}
\begin{figure*}
    \centering
       \includegraphics[width=1\linewidth]{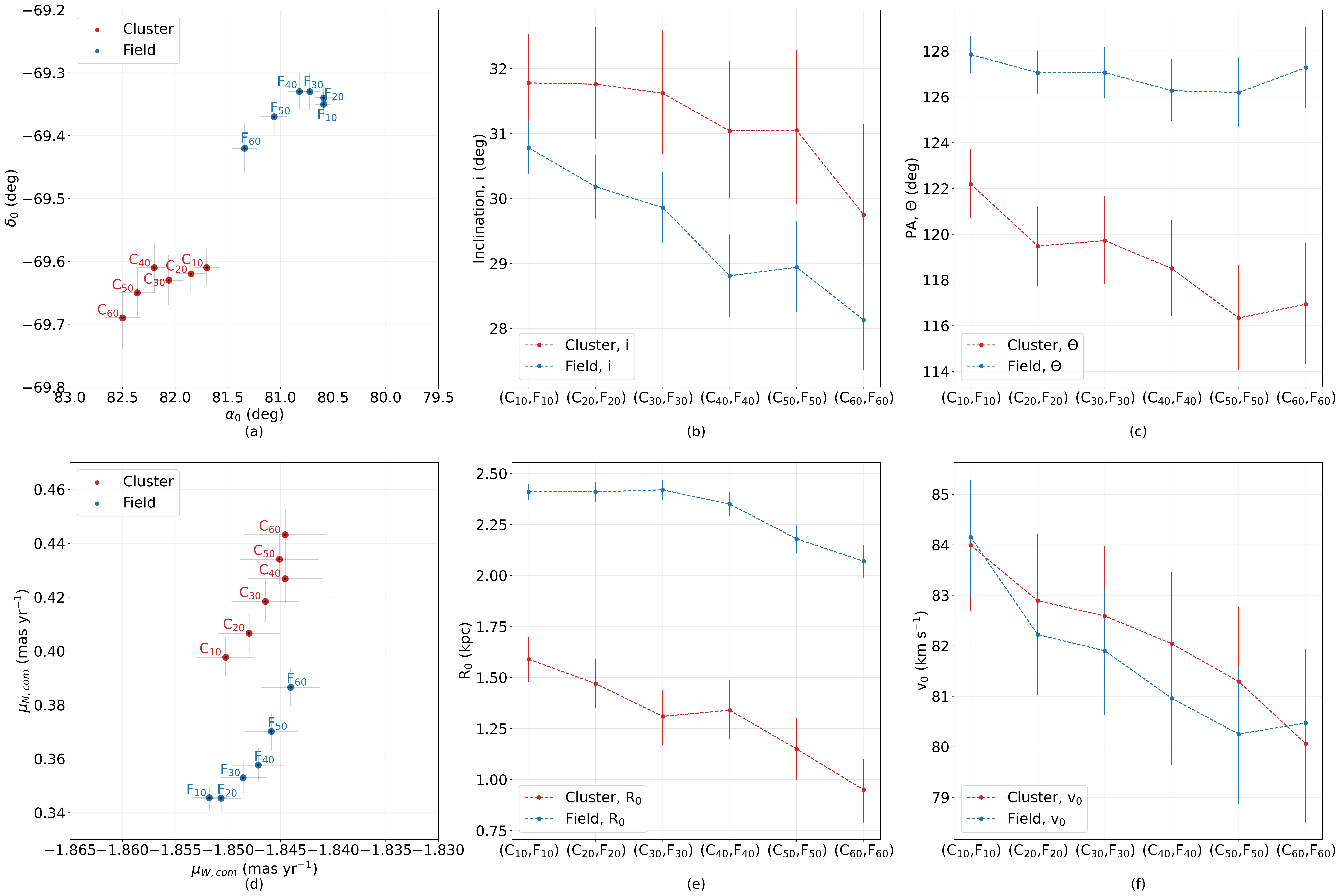}
       \caption{The variation of kinematic parameters based on cluster richness is shown here. Estimated parameters for cluster groups (C$_{10}$ to C$_{60}$) and nearby field groups (F$_{10}$ to F$_{60}$) are shown in red and blue colors from Panel (a) to (f). (a) Variation in ($\alpha_0$, $\delta_0$); (b) Variation in $i$; (c) Variation in $\Theta$; (d) Variation in ($\mu_{W,com}$, $\mu_{N,com}$); (e) Variation in $R_0$; (f) Variation in $v_0$.}
       \label{fig:six_richness} 
\end{figure*}

Figure \ref{fig:six_richness} shows the variation of the kinematic parameters corresponding to different cluster and field groups based on cluster richness. We note an offset of dynamic centers from poor to rich clusters in the increasing RA directions for both clusters and field populations. Meanwhile, the shift of COM PM for poor to rich clusters is seen increasing in North and West directions. Similarly, estimates for $\Phi$, $\Theta$, $R_0$, and $v_0$ show a decrease from C/F${10}$ to C/F${60}$ groups (as seen in Table \ref{tab5_paraemters}). However, a notable deviation becomes apparent in the kinematic parameters when considering entire cluster groups compared to field groups. This once again indicates that the kinematic model for clusters and the field population indeed manifests distinct kinematic properties. We also note that the cluster richness groups do not have similar spatial coverage. Overall, we do detect variation in the COM PM, dynamic center, $i$, and $R_0$, which may partially be due to the spatial coverage, and the variation in $\Theta$ is statistically insignificant.

\subsection{Impact of the LMC's spatial coverage on kinematic properties}\label{subsec_coverage}
Table \ref{tab6_paraemters} shows the estimated kinematic parameters for the data sets based on the spatial coverage of the LMC. The circular regions of radii increasing from 2$^\circ$ to 7$^\circ$ with a step size of 0.5$^\circ$ from the kinematic center (from the primary model, see Table \ref{tab2_paraemters}) of the LMC were used to perform the modeling.

\begin{table}
    \centering
    \caption{The kinematic best-fitting parameters obtained for the LMC based on the spatial coverage of the  LMC with clusters and nearby field regions are tabulated below.}
    \label{tab6_paraemters}
\renewcommand{\arraystretch}{1.5}
    \begin{tabular}{l@{\hspace{0.65em}}c@{\hspace{0.65em}}c@{\hspace{0.65em}}c@{\hspace{0.65em}}c@{\hspace{0.65em}}c@{\hspace{0.65em}}c@{\hspace{0.65em}}c@{\hspace{0.65em}}c@{\hspace{0.65em}}c@{\hspace{0.65em}}}
        \hline\hline
        \textbf{Data} & \textbf{$i$} & \textbf{$\Theta$} & \textbf{$\alpha_0$}& \textbf{$\delta_0$}& \textbf{$v_t$}& \textbf{$\theta_t$}& \textbf{$R_0$}& \textbf{$\eta$}& \textbf{$v_0$} \\
         & (deg) & (deg) & (deg) & (deg) & (km s$^{-1}$) & (deg) & (kpc) &  & (km s$^{-1}$) \\
        \hline
        C$_{r [{2.0}]}$& 49.93$^{\hspace{0.01cm}\scaleto{+1.36}{4.7pt}}_{\hspace{0.01cm}\scaleto{-1.41}{4.7pt}}$ & 104.79$^{\hspace{0.01cm}\scaleto{+1.64}{4.7pt}}_{\hspace{0.01cm}\scaleto{-1.63}{4.7pt}}$ & 82.45$^{\hspace{0.01cm}\scaleto{+0.3}{4.7pt}}_{\hspace{0.01cm}\scaleto{-0.29}{4.7pt}}$ & -69.89$^{\hspace{0.01cm}\scaleto{+0.05}{4.7pt}}_{\hspace{0.01cm}\scaleto{-0.05}{4.7pt}}$ & 457.7$^{\hspace{0.01cm}\scaleto{+1.59}{4.7pt}}_{\hspace{0.01cm}\scaleto{-1.56}{4.7pt}}$ & 166.76$^{\hspace{0.01cm}\scaleto{+0.53}{4.7pt}}_{\hspace{0.01cm}\scaleto{-0.56}{4.7pt}}$ & 1.41$^{\hspace{0.01cm}\scaleto{+0.11}{4.7pt}}_{\hspace{0.01cm}\scaleto{-0.12}{4.7pt}}$ & 3.38$^{\hspace{0.01cm}\scaleto{+0.99}{4.7pt}}_{\hspace{0.01cm}\scaleto{-0.67}{4.7pt}}$ & 75.84$^{\hspace{0.01cm}\scaleto{+1.78}{4.7pt}}_{\hspace{0.01cm}\scaleto{-1.81}{4.7pt}}$ \\
        F$_{r [{2.0}]}$& 42.89$^{\hspace{0.01cm}\scaleto{+0.77}{4.7pt}}_{\hspace{0.01cm}\scaleto{-0.8}{4.7pt}}$ & 110.07$^{\hspace{0.01cm}\scaleto{+1.06}{4.7pt}}_{\hspace{0.01cm}\scaleto{-1.07}{4.7pt}}$ & 81.58$^{\hspace{0.01cm}\scaleto{+0.13}{4.7pt}}_{\hspace{0.01cm}\scaleto{-0.13}{4.7pt}}$ & -69.86$^{\hspace{0.01cm}\scaleto{+0.03}{4.7pt}}_{\hspace{0.01cm}\scaleto{-0.03}{4.7pt}}$ & 455.62$^{\hspace{0.01cm}\scaleto{+0.82}{4.7pt}}_{\hspace{0.01cm}\scaleto{-0.76}{4.7pt}}$ & 167.66$^{\hspace{0.01cm}\scaleto{+0.24}{4.7pt}}_{\hspace{0.01cm}\scaleto{-0.24}{4.7pt}}$ & 1.49$^{\hspace{0.01cm}\scaleto{+0.09}{4.7pt}}_{\hspace{0.01cm}\scaleto{-0.08}{4.7pt}}$ & 1.57$^{\hspace{0.01cm}\scaleto{+0.13}{4.7pt}}_{\hspace{0.01cm}\scaleto{-0.12}{4.7pt}}$ & 77.5$^{\hspace{0.01cm}\scaleto{+1.76}{4.7pt}}_{\hspace{0.01cm}\scaleto{-1.72}{4.7pt}}$ \\
        \hline
        C$_{r [{2.5}]}$& 44.38$^{\hspace{0.01cm}\scaleto{+1.13}{4.7pt}}_{\hspace{0.01cm}\scaleto{-1.17}{4.7pt}}$ & 106.01$^{\hspace{0.01cm}\scaleto{+1.53}{4.7pt}}_{\hspace{0.01cm}\scaleto{-1.5}{4.7pt}}$ & 82.4$^{\hspace{0.01cm}\scaleto{+0.21}{4.7pt}}_{\hspace{0.01cm}\scaleto{-0.21}{4.7pt}}$ & -69.78$^{\hspace{0.01cm}\scaleto{+0.04}{4.7pt}}_{\hspace{0.01cm}\scaleto{-0.04}{4.7pt}}$ & 452.18$^{\hspace{0.01cm}\scaleto{+0.96}{4.7pt}}_{\hspace{0.01cm}\scaleto{-0.96}{4.7pt}}$ & 166.59$^{\hspace{0.01cm}\scaleto{+0.38}{4.7pt}}_{\hspace{0.01cm}\scaleto{-0.39}{4.7pt}}$ & 1.38$^{\hspace{0.01cm}\scaleto{+0.11}{4.7pt}}_{\hspace{0.01cm}\scaleto{-0.11}{4.7pt}}$ & 2.69$^{\hspace{0.01cm}\scaleto{+0.52}{4.7pt}}_{\hspace{0.01cm}\scaleto{-0.41}{4.7pt}}$ & 76.94$^{\hspace{0.01cm}\scaleto{+1.71}{4.7pt}}_{\hspace{0.01cm}\scaleto{-1.62}{4.7pt}}$ \\
        F$_{r [{2.5}]}$& 37.97$^{\hspace{0.01cm}\scaleto{+0.64}{4.7pt}}_{\hspace{0.01cm}\scaleto{-0.66}{4.7pt}}$ & 115.03$^{\hspace{0.01cm}\scaleto{+0.96}{4.7pt}}_{\hspace{0.01cm}\scaleto{-0.96}{4.7pt}}$ & 81.12$^{\hspace{0.01cm}\scaleto{+0.12}{4.7pt}}_{\hspace{0.01cm}\scaleto{-0.12}{4.7pt}}$ & -69.79$^{\hspace{0.01cm}\scaleto{+0.03}{4.7pt}}_{\hspace{0.01cm}\scaleto{-0.03}{4.7pt}}$ & 453.23$^{\hspace{0.01cm}\scaleto{+0.69}{4.7pt}}_{\hspace{0.01cm}\scaleto{-0.68}{4.7pt}}$ & 168.55$^{\hspace{0.01cm}\scaleto{+0.21}{4.7pt}}_{\hspace{0.01cm}\scaleto{-0.21}{4.7pt}}$ & 1.87$^{\hspace{0.01cm}\scaleto{+0.07}{4.7pt}}_{\hspace{0.01cm}\scaleto{-0.07}{4.7pt}}$ & 1.93$^{\hspace{0.01cm}\scaleto{+0.15}{4.7pt}}_{\hspace{0.01cm}\scaleto{-0.13}{4.7pt}}$ & 80.62$^{\hspace{0.01cm}\scaleto{+1.62}{4.7pt}}_{\hspace{0.01cm}\scaleto{-1.64}{4.7pt}}$ \\
        \hline
        C$_{r [{3.0}]}$& 39.51$^{\hspace{0.01cm}\scaleto{+1.07}{4.7pt}}_{\hspace{0.01cm}\scaleto{-1.08}{4.7pt}}$ & 109.7$^{\hspace{0.01cm}\scaleto{+1.53}{4.7pt}}_{\hspace{0.01cm}\scaleto{-1.55}{4.7pt}}$ & 82.16$^{\hspace{0.01cm}\scaleto{+0.15}{4.7pt}}_{\hspace{0.01cm}\scaleto{-0.15}{4.7pt}}$ & -69.71$^{\hspace{0.01cm}\scaleto{+0.03}{4.7pt}}_{\hspace{0.01cm}\scaleto{-0.03}{4.7pt}}$ & 448.62$^{\hspace{0.01cm}\scaleto{+0.76}{4.7pt}}_{\hspace{0.01cm}\scaleto{-0.78}{4.7pt}}$ & 167.02$^{\hspace{0.01cm}\scaleto{+0.27}{4.7pt}}_{\hspace{0.01cm}\scaleto{-0.27}{4.7pt}}$ & 1.32$^{\hspace{0.01cm}\scaleto{+0.11}{4.7pt}}_{\hspace{0.01cm}\scaleto{-0.11}{4.7pt}}$ & 2.19$^{\hspace{0.01cm}\scaleto{+0.33}{4.7pt}}_{\hspace{0.01cm}\scaleto{-0.27}{4.7pt}}$ & 77.67$^{\hspace{0.01cm}\scaleto{+1.61}{4.7pt}}_{\hspace{0.01cm}\scaleto{-1.58}{4.7pt}}$ \\
        F$_{r [{3.0}]}$& 33.93$^{\hspace{0.01cm}\scaleto{+0.54}{4.7pt}}_{\hspace{0.01cm}\scaleto{-0.56}{4.7pt}}$ & 119.91$^{\hspace{0.01cm}\scaleto{+0.92}{4.7pt}}_{\hspace{0.01cm}\scaleto{-0.91}{4.7pt}}$ & 80.74$^{\hspace{0.01cm}\scaleto{+0.09}{4.7pt}}_{\hspace{0.01cm}\scaleto{-0.1}{4.7pt}}$ & -69.63$^{\hspace{0.01cm}\scaleto{+0.03}{4.7pt}}_{\hspace{0.01cm}\scaleto{-0.03}{4.7pt}}$ & 448.94$^{\hspace{0.01cm}\scaleto{+0.49}{4.7pt}}_{\hspace{0.01cm}\scaleto{-0.49}{4.7pt}}$ & 169.2$^{\hspace{0.01cm}\scaleto{+0.17}{4.7pt}}_{\hspace{0.01cm}\scaleto{-0.17}{4.7pt}}$ & 2.04$^{\hspace{0.01cm}\scaleto{+0.06}{4.7pt}}_{\hspace{0.01cm}\scaleto{-0.06}{4.7pt}}$ & 2.03$^{\hspace{0.01cm}\scaleto{+0.14}{4.7pt}}_{\hspace{0.01cm}\scaleto{-0.12}{4.7pt}}$ & 81.53$^{\hspace{0.01cm}\scaleto{+1.55}{4.7pt}}_{\hspace{0.01cm}\scaleto{-1.58}{4.7pt}}$ \\
        \hline
        C$_{r [{3.5}]}$& 37.35$^{\hspace{0.01cm}\scaleto{+0.94}{4.7pt}}_{\hspace{0.01cm}\scaleto{-0.99}{4.7pt}}$ & 113.34$^{\hspace{0.01cm}\scaleto{+1.49}{4.7pt}}_{\hspace{0.01cm}\scaleto{-1.54}{4.7pt}}$ & 81.98$^{\hspace{0.01cm}\scaleto{+0.13}{4.7pt}}_{\hspace{0.01cm}\scaleto{-0.13}{4.7pt}}$ & -69.71$^{\hspace{0.01cm}\scaleto{+0.03}{4.7pt}}_{\hspace{0.01cm}\scaleto{-0.03}{4.7pt}}$ & 447.66$^{\hspace{0.01cm}\scaleto{+0.71}{4.7pt}}_{\hspace{0.01cm}\scaleto{-0.68}{4.7pt}}$ & 167.43$^{\hspace{0.01cm}\scaleto{+0.23}{4.7pt}}_{\hspace{0.01cm}\scaleto{-0.23}{4.7pt}}$ & 1.32$^{\hspace{0.01cm}\scaleto{+0.1}{4.7pt}}_{\hspace{0.01cm}\scaleto{-0.11}{4.7pt}}$ & 2.09$^{\hspace{0.01cm}\scaleto{+0.27}{4.7pt}}_{\hspace{0.01cm}\scaleto{-0.23}{4.7pt}}$ & 78.5$^{\hspace{0.01cm}\scaleto{+1.53}{4.7pt}}_{\hspace{0.01cm}\scaleto{-1.45}{4.7pt}}$ \\
        F$_{r [{3.5}]}$& 32.35$^{\hspace{0.01cm}\scaleto{+0.49}{4.7pt}}_{\hspace{0.01cm}\scaleto{-0.5}{4.7pt}}$ & 122.78$^{\hspace{0.01cm}\scaleto{+0.86}{4.7pt}}_{\hspace{0.01cm}\scaleto{-0.86}{4.7pt}}$ & 80.59$^{\hspace{0.01cm}\scaleto{+0.08}{4.7pt}}_{\hspace{0.01cm}\scaleto{-0.08}{4.7pt}}$ & -69.5$^{\hspace{0.01cm}\scaleto{+0.02}{4.7pt}}_{\hspace{0.01cm}\scaleto{-0.02}{4.7pt}}$ & 445.93$^{\hspace{0.01cm}\scaleto{+0.41}{4.7pt}}_{\hspace{0.01cm}\scaleto{-0.41}{4.7pt}}$ & 169.45$^{\hspace{0.01cm}\scaleto{+0.15}{4.7pt}}_{\hspace{0.01cm}\scaleto{-0.15}{4.7pt}}$ & 2.15$^{\hspace{0.01cm}\scaleto{+0.05}{4.7pt}}_{\hspace{0.01cm}\scaleto{-0.05}{4.7pt}}$ & 2.13$^{\hspace{0.01cm}\scaleto{+0.14}{4.7pt}}_{\hspace{0.01cm}\scaleto{-0.13}{4.7pt}}$ & 81.67$^{\hspace{0.01cm}\scaleto{+1.51}{4.7pt}}_{\hspace{0.01cm}\scaleto{-1.47}{4.7pt}}$ \\
        \hline
        C$_{r [{4.0}]}$& 35.03$^{\hspace{0.01cm}\scaleto{+0.9}{4.7pt}}_{\hspace{0.01cm}\scaleto{-0.96}{4.7pt}}$ & 115.7$^{\hspace{0.01cm}\scaleto{+1.56}{4.7pt}}_{\hspace{0.01cm}\scaleto{-1.57}{4.7pt}}$ & 81.87$^{\hspace{0.01cm}\scaleto{+0.13}{4.7pt}}_{\hspace{0.01cm}\scaleto{-0.13}{4.7pt}}$ & -69.69$^{\hspace{0.01cm}\scaleto{+0.03}{4.7pt}}_{\hspace{0.01cm}\scaleto{-0.03}{4.7pt}}$ & 446.66$^{\hspace{0.01cm}\scaleto{+0.66}{4.7pt}}_{\hspace{0.01cm}\scaleto{-0.66}{4.7pt}}$ & 167.61$^{\hspace{0.01cm}\scaleto{+0.23}{4.7pt}}_{\hspace{0.01cm}\scaleto{-0.22}{4.7pt}}$ & 1.35$^{\hspace{0.01cm}\scaleto{+0.1}{4.7pt}}_{\hspace{0.01cm}\scaleto{-0.11}{4.7pt}}$ & 1.95$^{\hspace{0.01cm}\scaleto{+0.22}{4.7pt}}_{\hspace{0.01cm}\scaleto{-0.2}{4.7pt}}$ & 79.76$^{\hspace{0.01cm}\scaleto{+1.51}{4.7pt}}_{\hspace{0.01cm}\scaleto{-1.48}{4.7pt}}$ \\
        F$_{r [{4.0}]}$& 30.64$^{\hspace{0.01cm}\scaleto{+0.46}{4.7pt}}_{\hspace{0.01cm}\scaleto{-0.49}{4.7pt}}$ & 125.48$^{\hspace{0.01cm}\scaleto{+0.85}{4.7pt}}_{\hspace{0.01cm}\scaleto{-0.87}{4.7pt}}$ & 80.39$^{\hspace{0.01cm}\scaleto{+0.08}{4.7pt}}_{\hspace{0.01cm}\scaleto{-0.08}{4.7pt}}$ & -69.38$^{\hspace{0.01cm}\scaleto{+0.02}{4.7pt}}_{\hspace{0.01cm}\scaleto{-0.02}{4.7pt}}$ & 443.67$^{\hspace{0.01cm}\scaleto{+0.42}{4.7pt}}_{\hspace{0.01cm}\scaleto{-0.41}{4.7pt}}$ & 169.82$^{\hspace{0.01cm}\scaleto{+0.15}{4.7pt}}_{\hspace{0.01cm}\scaleto{-0.15}{4.7pt}}$ & 2.37$^{\hspace{0.01cm}\scaleto{+0.05}{4.7pt}}_{\hspace{0.01cm}\scaleto{-0.05}{4.7pt}}$ & 2.29$^{\hspace{0.01cm}\scaleto{+0.14}{4.7pt}}_{\hspace{0.01cm}\scaleto{-0.13}{4.7pt}}$ & 83.74$^{\hspace{0.01cm}\scaleto{+1.49}{4.7pt}}_{\hspace{0.01cm}\scaleto{-1.42}{4.7pt}}$ \\
        \hline
        C$_{r [{4.5}]}$& 33.33$^{\hspace{0.01cm}\scaleto{+0.88}{4.7pt}}_{\hspace{0.01cm}\scaleto{-0.91}{4.7pt}}$ & 118.37$^{\hspace{0.01cm}\scaleto{+1.48}{4.7pt}}_{\hspace{0.01cm}\scaleto{-1.54}{4.7pt}}$ & 81.63$^{\hspace{0.01cm}\scaleto{+0.13}{4.7pt}}_{\hspace{0.01cm}\scaleto{-0.13}{4.7pt}}$ & -69.61$^{\hspace{0.01cm}\scaleto{+0.03}{4.7pt}}_{\hspace{0.01cm}\scaleto{-0.03}{4.7pt}}$ & 444.86$^{\hspace{0.01cm}\scaleto{+0.63}{4.7pt}}_{\hspace{0.01cm}\scaleto{-0.64}{4.7pt}}$ & 167.99$^{\hspace{0.01cm}\scaleto{+0.23}{4.7pt}}_{\hspace{0.01cm}\scaleto{-0.23}{4.7pt}}$ & 1.55$^{\hspace{0.01cm}\scaleto{+0.11}{4.7pt}}_{\hspace{0.01cm}\scaleto{-0.11}{4.7pt}}$ & 1.94$^{\hspace{0.01cm}\scaleto{+0.19}{4.7pt}}_{\hspace{0.01cm}\scaleto{-0.17}{4.7pt}}$ & 82.72$^{\hspace{0.01cm}\scaleto{+1.39}{4.7pt}}_{\hspace{0.01cm}\scaleto{-1.34}{4.7pt}}$ \\
        F$_{r [{4.5}]}$& 30.43$^{\hspace{0.01cm}\scaleto{+0.45}{4.7pt}}_{\hspace{0.01cm}\scaleto{-0.45}{4.7pt}}$ & 126.41$^{\hspace{0.01cm}\scaleto{+0.87}{4.7pt}}_{\hspace{0.01cm}\scaleto{-0.85}{4.7pt}}$ & 80.44$^{\hspace{0.01cm}\scaleto{+0.08}{4.7pt}}_{\hspace{0.01cm}\scaleto{-0.08}{4.7pt}}$ & -69.35$^{\hspace{0.01cm}\scaleto{+0.02}{4.7pt}}_{\hspace{0.01cm}\scaleto{-0.02}{4.7pt}}$ & 442.91$^{\hspace{0.01cm}\scaleto{+0.38}{4.7pt}}_{\hspace{0.01cm}\scaleto{-0.39}{4.7pt}}$ & 169.73$^{\hspace{0.01cm}\scaleto{+0.14}{4.7pt}}_{\hspace{0.01cm}\scaleto{-0.15}{4.7pt}}$ & 2.43$^{\hspace{0.01cm}\scaleto{+0.05}{4.7pt}}_{\hspace{0.01cm}\scaleto{-0.05}{4.7pt}}$ & 2.29$^{\hspace{0.01cm}\scaleto{+0.14}{4.7pt}}_{\hspace{0.01cm}\scaleto{-0.13}{4.7pt}}$ & 84.77$^{\hspace{0.01cm}\scaleto{+1.43}{4.7pt}}_{\hspace{0.01cm}\scaleto{-1.36}{4.7pt}}$ \\
        \hline
        C$_{r [{5.0}]}$& 32.34$^{\hspace{0.01cm}\scaleto{+0.86}{4.7pt}}_{\hspace{0.01cm}\scaleto{-0.89}{4.7pt}}$ & 119.75$^{\hspace{0.01cm}\scaleto{+1.53}{4.7pt}}_{\hspace{0.01cm}\scaleto{-1.5}{4.7pt}}$ & 81.61$^{\hspace{0.01cm}\scaleto{+0.13}{4.7pt}}_{\hspace{0.01cm}\scaleto{-0.13}{4.7pt}}$ & -69.6$^{\hspace{0.01cm}\scaleto{+0.03}{4.7pt}}_{\hspace{0.01cm}\scaleto{-0.03}{4.7pt}}$ & 444.66$^{\hspace{0.01cm}\scaleto{+0.62}{4.7pt}}_{\hspace{0.01cm}\scaleto{-0.6}{4.7pt}}$ & 168.04$^{\hspace{0.01cm}\scaleto{+0.22}{4.7pt}}_{\hspace{0.01cm}\scaleto{-0.22}{4.7pt}}$ & 1.6$^{\hspace{0.01cm}\scaleto{+0.11}{4.7pt}}_{\hspace{0.01cm}\scaleto{-0.11}{4.7pt}}$ & 1.9$^{\hspace{0.01cm}\scaleto{+0.18}{4.7pt}}_{\hspace{0.01cm}\scaleto{-0.16}{4.7pt}}$ & 83.67$^{\hspace{0.01cm}\scaleto{+1.39}{4.7pt}}_{\hspace{0.01cm}\scaleto{-1.35}{4.7pt}}$ \\
        F$_{r [{5.0}]}$& 30.39$^{\hspace{0.01cm}\scaleto{+0.42}{4.7pt}}_{\hspace{0.01cm}\scaleto{-0.44}{4.7pt}}$ & 127.44$^{\hspace{0.01cm}\scaleto{+0.83}{4.7pt}}_{\hspace{0.01cm}\scaleto{-0.85}{4.7pt}}$ & 80.5$^{\hspace{0.01cm}\scaleto{+0.08}{4.7pt}}_{\hspace{0.01cm}\scaleto{-0.08}{4.7pt}}$ & -69.34$^{\hspace{0.01cm}\scaleto{+0.02}{4.7pt}}_{\hspace{0.01cm}\scaleto{-0.02}{4.7pt}}$ & 442.65$^{\hspace{0.01cm}\scaleto{+0.36}{4.7pt}}_{\hspace{0.01cm}\scaleto{-0.36}{4.7pt}}$ & 169.61$^{\hspace{0.01cm}\scaleto{+0.14}{4.7pt}}_{\hspace{0.01cm}\scaleto{-0.14}{4.7pt}}$ & 2.45$^{\hspace{0.01cm}\scaleto{+0.05}{4.7pt}}_{\hspace{0.01cm}\scaleto{-0.05}{4.7pt}}$ & 2.27$^{\hspace{0.01cm}\scaleto{+0.13}{4.7pt}}_{\hspace{0.01cm}\scaleto{-0.13}{4.7pt}}$ & 85.32$^{\hspace{0.01cm}\scaleto{+1.32}{4.7pt}}_{\hspace{0.01cm}\scaleto{-1.31}{4.7pt}}$ \\
        \hline
        C$_{r [{5.5}]}$& 31.68$^{\hspace{0.01cm}\scaleto{+0.81}{4.7pt}}_{\hspace{0.01cm}\scaleto{-0.85}{4.7pt}}$ & 120.95$^{\hspace{0.01cm}\scaleto{+1.52}{4.7pt}}_{\hspace{0.01cm}\scaleto{-1.52}{4.7pt}}$ & 81.64$^{\hspace{0.01cm}\scaleto{+0.12}{4.7pt}}_{\hspace{0.01cm}\scaleto{-0.13}{4.7pt}}$ & -69.61$^{\hspace{0.01cm}\scaleto{+0.03}{4.7pt}}_{\hspace{0.01cm}\scaleto{-0.03}{4.7pt}}$ & 444.8$^{\hspace{0.01cm}\scaleto{+0.56}{4.7pt}}_{\hspace{0.01cm}\scaleto{-0.57}{4.7pt}}$ & 168.0$^{\hspace{0.01cm}\scaleto{+0.22}{4.7pt}}_{\hspace{0.01cm}\scaleto{-0.22}{4.7pt}}$ & 1.58$^{\hspace{0.01cm}\scaleto{+0.1}{4.7pt}}_{\hspace{0.01cm}\scaleto{-0.1}{4.7pt}}$ & 1.86$^{\hspace{0.01cm}\scaleto{+0.17}{4.7pt}}_{\hspace{0.01cm}\scaleto{-0.15}{4.7pt}}$ & 83.55$^{\hspace{0.01cm}\scaleto{+1.36}{4.7pt}}_{\hspace{0.01cm}\scaleto{-1.29}{4.7pt}}$ \\
        F$_{r [{5.5}]}$& 30.58$^{\hspace{0.01cm}\scaleto{+0.41}{4.7pt}}_{\hspace{0.01cm}\scaleto{-0.42}{4.7pt}}$ & 127.76$^{\hspace{0.01cm}\scaleto{+0.8}{4.7pt}}_{\hspace{0.01cm}\scaleto{-0.82}{4.7pt}}$ & 80.54$^{\hspace{0.01cm}\scaleto{+0.08}{4.7pt}}_{\hspace{0.01cm}\scaleto{-0.08}{4.7pt}}$ & -69.35$^{\hspace{0.01cm}\scaleto{+0.02}{4.7pt}}_{\hspace{0.01cm}\scaleto{-0.02}{4.7pt}}$ & 442.79$^{\hspace{0.01cm}\scaleto{+0.36}{4.7pt}}_{\hspace{0.01cm}\scaleto{-0.37}{4.7pt}}$ & 169.52$^{\hspace{0.01cm}\scaleto{+0.14}{4.7pt}}_{\hspace{0.01cm}\scaleto{-0.14}{4.7pt}}$ & 2.45$^{\hspace{0.01cm}\scaleto{+0.04}{4.7pt}}_{\hspace{0.01cm}\scaleto{-0.04}{4.7pt}}$ & 2.28$^{\hspace{0.01cm}\scaleto{+0.13}{4.7pt}}_{\hspace{0.01cm}\scaleto{-0.13}{4.7pt}}$ & 85.28$^{\hspace{0.01cm}\scaleto{+1.32}{4.7pt}}_{\hspace{0.01cm}\scaleto{-1.22}{4.7pt}}$ \\
        \hline
        C$_{r [{6.0}]}$& 31.35$^{\hspace{0.01cm}\scaleto{+0.76}{4.7pt}}_{\hspace{0.01cm}\scaleto{-0.78}{4.7pt}}$ & 121.69$^{\hspace{0.01cm}\scaleto{+1.52}{4.7pt}}_{\hspace{0.01cm}\scaleto{-1.53}{4.7pt}}$ & 81.66$^{\hspace{0.01cm}\scaleto{+0.13}{4.7pt}}_{\hspace{0.01cm}\scaleto{-0.13}{4.7pt}}$ & -69.61$^{\hspace{0.01cm}\scaleto{+0.03}{4.7pt}}_{\hspace{0.01cm}\scaleto{-0.03}{4.7pt}}$ & 444.79$^{\hspace{0.01cm}\scaleto{+0.53}{4.7pt}}_{\hspace{0.01cm}\scaleto{-0.55}{4.7pt}}$ & 167.95$^{\hspace{0.01cm}\scaleto{+0.21}{4.7pt}}_{\hspace{0.01cm}\scaleto{-0.22}{4.7pt}}$ & 1.58$^{\hspace{0.01cm}\scaleto{+0.1}{4.7pt}}_{\hspace{0.01cm}\scaleto{-0.11}{4.7pt}}$ & 1.84$^{\hspace{0.01cm}\scaleto{+0.17}{4.7pt}}_{\hspace{0.01cm}\scaleto{-0.15}{4.7pt}}$ & 83.68$^{\hspace{0.01cm}\scaleto{+1.32}{4.7pt}}_{\hspace{0.01cm}\scaleto{-1.25}{4.7pt}}$ \\
        F$_{r [{6.0}]}$& 30.66$^{\hspace{0.01cm}\scaleto{+0.4}{4.7pt}}_{\hspace{0.01cm}\scaleto{-0.41}{4.7pt}}$ & 127.96$^{\hspace{0.01cm}\scaleto{+0.78}{4.7pt}}_{\hspace{0.01cm}\scaleto{-0.8}{4.7pt}}$ & 80.57$^{\hspace{0.01cm}\scaleto{+0.08}{4.7pt}}_{\hspace{0.01cm}\scaleto{-0.07}{4.7pt}}$ & -69.34$^{\hspace{0.01cm}\scaleto{+0.02}{4.7pt}}_{\hspace{0.01cm}\scaleto{-0.02}{4.7pt}}$ & 442.68$^{\hspace{0.01cm}\scaleto{+0.34}{4.7pt}}_{\hspace{0.01cm}\scaleto{-0.35}{4.7pt}}$ & 169.47$^{\hspace{0.01cm}\scaleto{+0.14}{4.7pt}}_{\hspace{0.01cm}\scaleto{-0.14}{4.7pt}}$ & 2.44$^{\hspace{0.01cm}\scaleto{+0.04}{4.7pt}}_{\hspace{0.01cm}\scaleto{-0.04}{4.7pt}}$ & 2.3$^{\hspace{0.01cm}\scaleto{+0.14}{4.7pt}}_{\hspace{0.01cm}\scaleto{-0.13}{4.7pt}}$ & 84.87$^{\hspace{0.01cm}\scaleto{+1.26}{4.7pt}}_{\hspace{0.01cm}\scaleto{-1.19}{4.7pt}}$ \\
        \hline
        C$_{r [{6.5}]}$& 31.61$^{\hspace{0.01cm}\scaleto{+0.75}{4.7pt}}_{\hspace{0.01cm}\scaleto{-0.77}{4.7pt}}$ & 121.87$^{\hspace{0.01cm}\scaleto{+1.54}{4.7pt}}_{\hspace{0.01cm}\scaleto{-1.54}{4.7pt}}$ & 81.65$^{\hspace{0.01cm}\scaleto{+0.12}{4.7pt}}_{\hspace{0.01cm}\scaleto{-0.12}{4.7pt}}$ & -69.6$^{\hspace{0.01cm}\scaleto{+0.03}{4.7pt}}_{\hspace{0.01cm}\scaleto{-0.03}{4.7pt}}$ & 444.59$^{\hspace{0.01cm}\scaleto{+0.54}{4.7pt}}_{\hspace{0.01cm}\scaleto{-0.53}{4.7pt}}$ & 167.96$^{\hspace{0.01cm}\scaleto{+0.21}{4.7pt}}_{\hspace{0.01cm}\scaleto{-0.21}{4.7pt}}$ & 1.57$^{\hspace{0.01cm}\scaleto{+0.1}{4.7pt}}_{\hspace{0.01cm}\scaleto{-0.11}{4.7pt}}$ & 1.87$^{\hspace{0.01cm}\scaleto{+0.17}{4.7pt}}_{\hspace{0.01cm}\scaleto{-0.16}{4.7pt}}$ & 83.35$^{\hspace{0.01cm}\scaleto{+1.3}{4.7pt}}_{\hspace{0.01cm}\scaleto{-1.24}{4.7pt}}$ \\
        F$_{r [{6.5}]}$& 30.72$^{\hspace{0.01cm}\scaleto{+0.39}{4.7pt}}_{\hspace{0.01cm}\scaleto{-0.4}{4.7pt}}$ & 127.99$^{\hspace{0.01cm}\scaleto{+0.79}{4.7pt}}_{\hspace{0.01cm}\scaleto{-0.8}{4.7pt}}$ & 80.57$^{\hspace{0.01cm}\scaleto{+0.07}{4.7pt}}_{\hspace{0.01cm}\scaleto{-0.07}{4.7pt}}$ & -69.34$^{\hspace{0.01cm}\scaleto{+0.02}{4.7pt}}_{\hspace{0.01cm}\scaleto{-0.02}{4.7pt}}$ & 442.66$^{\hspace{0.01cm}\scaleto{+0.34}{4.7pt}}_{\hspace{0.01cm}\scaleto{-0.33}{4.7pt}}$ & 169.47$^{\hspace{0.01cm}\scaleto{+0.13}{4.7pt}}_{\hspace{0.01cm}\scaleto{-0.13}{4.7pt}}$ & 2.44$^{\hspace{0.01cm}\scaleto{+0.04}{4.7pt}}_{\hspace{0.01cm}\scaleto{-0.04}{4.7pt}}$ & 2.34$^{\hspace{0.01cm}\scaleto{+0.14}{4.7pt}}_{\hspace{0.01cm}\scaleto{-0.13}{4.7pt}}$ & 84.45$^{\hspace{0.01cm}\scaleto{+1.14}{4.7pt}}_{\hspace{0.01cm}\scaleto{-1.15}{4.7pt}}$ \\
        \hline
        C$_{r [{7.0}]}$& 31.54$^{\hspace{0.01cm}\scaleto{+0.73}{4.7pt}}_{\hspace{0.01cm}\scaleto{-0.74}{4.7pt}}$ & 122.31$^{\hspace{0.01cm}\scaleto{+1.47}{4.7pt}}_{\hspace{0.01cm}\scaleto{-1.52}{4.7pt}}$ & 81.69$^{\hspace{0.01cm}\scaleto{+0.12}{4.7pt}}_{\hspace{0.01cm}\scaleto{-0.12}{4.7pt}}$ & -69.6$^{\hspace{0.01cm}\scaleto{+0.03}{4.7pt}}_{\hspace{0.01cm}\scaleto{-0.03}{4.7pt}}$ & 444.55$^{\hspace{0.01cm}\scaleto{+0.54}{4.7pt}}_{\hspace{0.01cm}\scaleto{-0.52}{4.7pt}}$ & 167.89$^{\hspace{0.01cm}\scaleto{+0.21}{4.7pt}}_{\hspace{0.01cm}\scaleto{-0.21}{4.7pt}}$ & 1.58$^{\hspace{0.01cm}\scaleto{+0.1}{4.7pt}}_{\hspace{0.01cm}\scaleto{-0.11}{4.7pt}}$ & 1.84$^{\hspace{0.01cm}\scaleto{+0.16}{4.7pt}}_{\hspace{0.01cm}\scaleto{-0.15}{4.7pt}}$ & 83.9$^{\hspace{0.01cm}\scaleto{+1.25}{4.7pt}}_{\hspace{0.01cm}\scaleto{-1.24}{4.7pt}}$ \\
        F$_{r [{7.0}]}$& 30.79$^{\hspace{0.01cm}\scaleto{+0.39}{4.7pt}}_{\hspace{0.01cm}\scaleto{-0.39}{4.7pt}}$ & 128.04$^{\hspace{0.01cm}\scaleto{+0.78}{4.7pt}}_{\hspace{0.01cm}\scaleto{-0.8}{4.7pt}}$ & 80.57$^{\hspace{0.01cm}\scaleto{+0.07}{4.7pt}}_{\hspace{0.01cm}\scaleto{-0.07}{4.7pt}}$ & -69.34$^{\hspace{0.01cm}\scaleto{+0.02}{4.7pt}}_{\hspace{0.01cm}\scaleto{-0.02}{4.7pt}}$ & 442.7$^{\hspace{0.01cm}\scaleto{+0.34}{4.7pt}}_{\hspace{0.01cm}\scaleto{-0.34}{4.7pt}}$ & 169.46$^{\hspace{0.01cm}\scaleto{+0.13}{4.7pt}}_{\hspace{0.01cm}\scaleto{-0.13}{4.7pt}}$ & 2.43$^{\hspace{0.01cm}\scaleto{+0.04}{4.7pt}}_{\hspace{0.01cm}\scaleto{-0.04}{4.7pt}}$ & 2.36$^{\hspace{0.01cm}\scaleto{+0.14}{4.7pt}}_{\hspace{0.01cm}\scaleto{-0.13}{4.7pt}}$ & 84.18$^{\hspace{0.01cm}\scaleto{+1.14}{4.7pt}}_{\hspace{0.01cm}\scaleto{-1.1}{4.7pt}}$ \\
        \hline
        \hline
    \end{tabular}
\end{table}
\begin{figure*}
    \centering
       \includegraphics[width=1\linewidth]{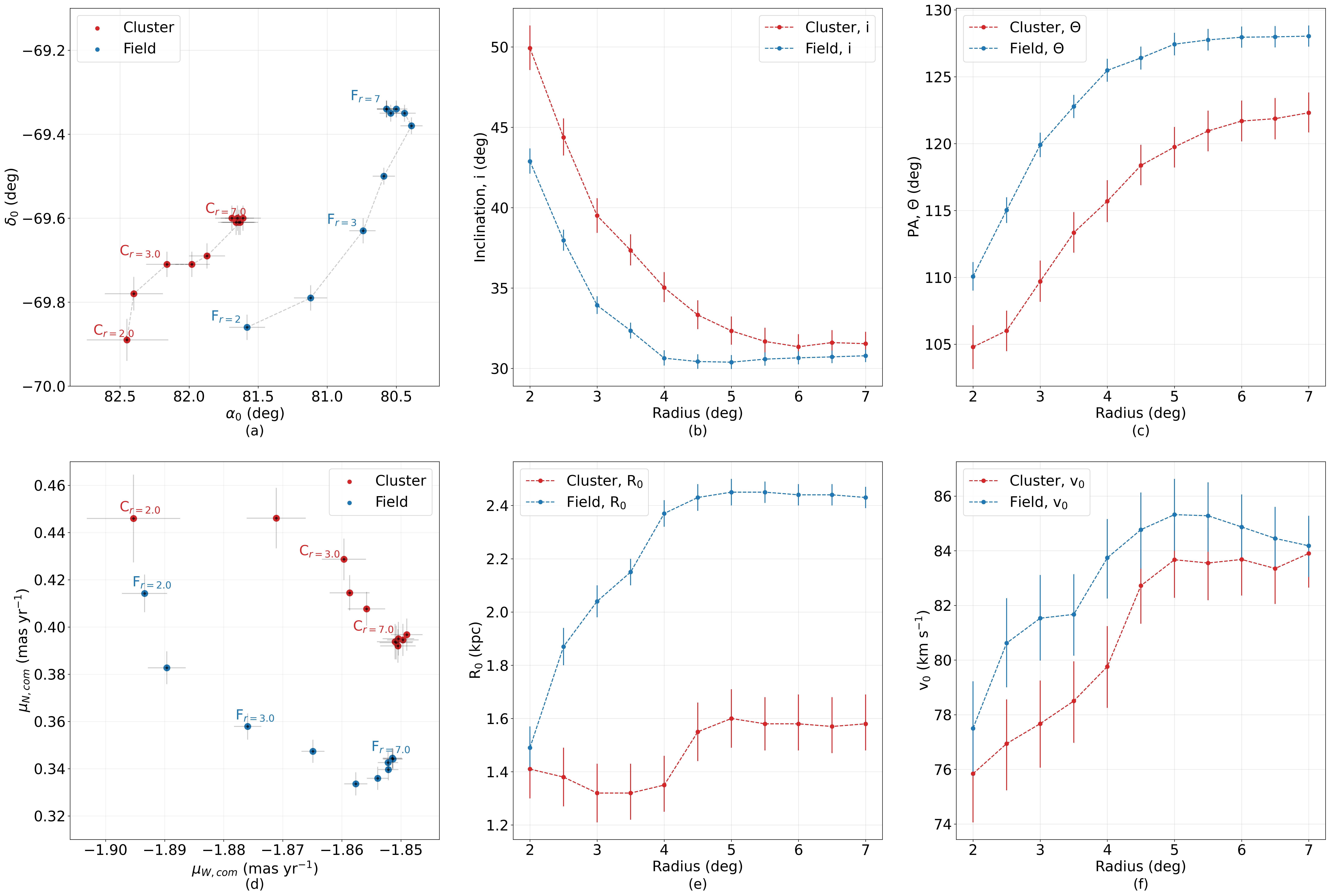}
       \caption{The variation of kinematic parameters based on the spatial coverage of the LMC is shown here. Estimated parameters for different spatial coverage from 2$^\circ$ to 7$^\circ$ with a step size of 0.5$^\circ$ from the center (based on the primary model) of the LMC are shown here. The corresponding cluster data sets (C$_{r=2}$ to C$_{r=7}$) and nearby field data sets (F$_{r=2}$ to F$_{r=7}$)are shown in red and blue colors from Panel (a) to (f). (a) Variation in ($\alpha_0$, $\delta_0$); (b) Variation in $i$; (c) Variation in $\Theta$; (d) Variation in ($\mu_{W,com}$, $\mu_{N,com}$); (e) Variation in $R_0$; (f) Variation in $v_0$.}
       \label{fig:six_r} 
\end{figure*}

Figure \ref{fig:six_r} shows the radial variation of the kinematic parameters in the case of clusters and nearby field regions. Within the radii considered in this study, there is a notable offset in the amplitude of PM, of the order of $\sim$ 0.06 mas yr$^{-1}$ in ($\mu_{W,com}$, $\mu_{N,com}$) between the inner and outer cluster/field regions.
Similarly, we note a positional offset of $\sim$ 0.4$^\circ$ for clusters and $\sim$ 0.6$^\circ$ for the field between the estimated values of $\alpha_0$ and $\delta_0$. The value of  $i$ steeply decreases from the inner to outer radii, from $\sim$ 50$^\circ$ to $\sim$ 32$^\circ$ in the case of clusters, and $\sim$ 43$^\circ$ to $\sim$ 31$^\circ$ in the case of field population. Meanwhile, the value of $\Theta$ increases from inner to outer radii, from $\sim$ 108$^\circ$ to $\sim$ 122$^\circ$ in the case of clusters, and $\sim$ 110$^\circ$ to $\sim$ 128$^\circ$ in the case of field population. The rotational parameters, $R_0$ and $v_0$ after 5$^\circ$ show convergence to the values estimated as in the primary model (see Table \ref{tab2_paraemters}). Therefore, the estimated parameters show a significant radial dependence for both clusters and field populations.

\subsection{Residual PM of the LMC: Cluster vs Field}

\begin{figure*}
    \centering
       \includegraphics[width=1\linewidth]{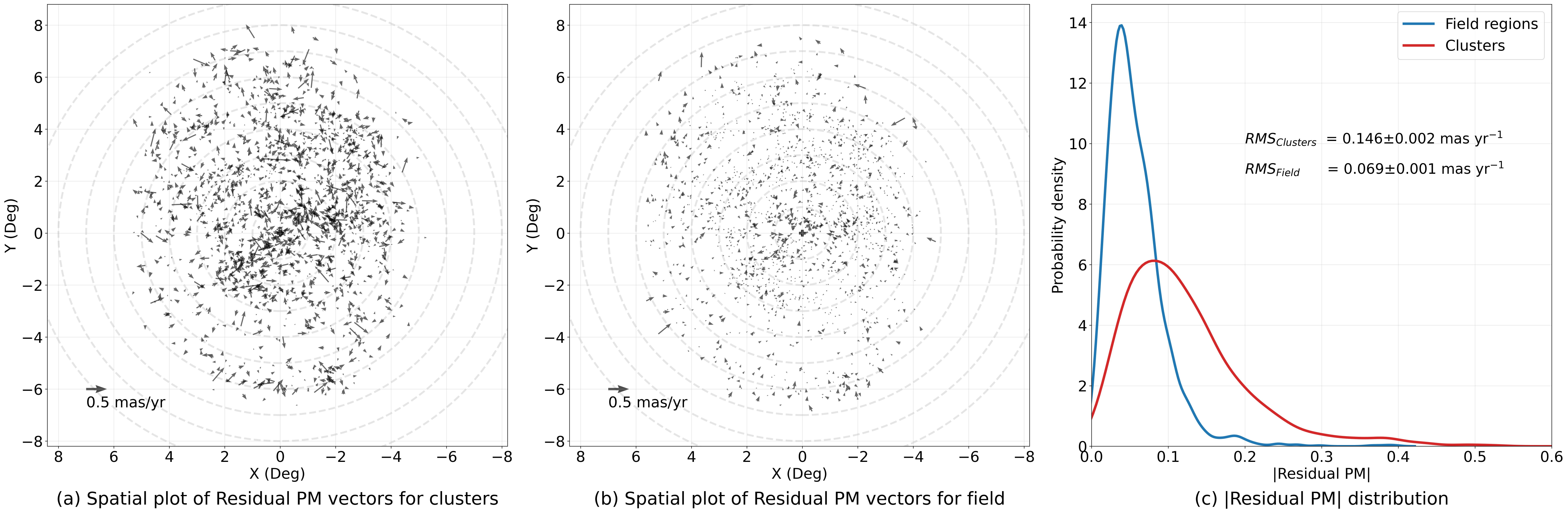}
       \caption{Residual PM of clusters and field population. (a) Spatial plot of residual PM vectors for the clusters; (b) Spatial distribution of residual PM vectors for the field population; (c) The Gaussian KDE showing the distribution of the $|$Residual PM$|$.}
       \label{fig:respm_plots} 
\end{figure*}

The residual PM value for the clusters and nearby field populations is found by subtracting the net modeled PM from the net observed PM values. The residual PM vectors for clusters and field regions are plotted in the X-Y plane as shown in Figure \ref{fig:respm_plots}, panels (a) and (b), respectively. Clusters show larger residual PM in the spatial plot when compared with the field population. The spatial residual plot for clusters shows relatively large residuals in the bar region and in the northern LMC, whereas such large residuals are not found in the corresponding plot for the field population. 
The residual PM amplitudes ($|$Residual PM$|$) of both the clusters and the nearby field are used to generate the probability distribution plot as in Figure \ref{fig:respm_plots}, panel (c) depicting the distribution of their values. We obtained an RMS value of 0.146$\pm$0.002 mas yr$^{-1}$ for clusters and 0.069$\pm$0.001 mas yr$^{-1}$ for the nearby field population. The RMS distribution of clusters shows a broader profile, while the field shows a narrower one.

\section{Discussion}\label{sec_5}
In this study, we performed the kinematic modeling of the LMC using 1705 star clusters and nearby field regions for the first time. We also created an extensive additional 46 models using two control samples (young MS stars and RC stars), different cluster age groups and nearby fields, cluster richness groups, and samples with varying spatial coverage of the  LMC. The clusters were cleaned using field star contamination (D24) and found less than 2\% of MW source contamination in each cluster while cross-matching with the MW/ LMC source catalog by J23. The median value of the PM is used for the study for both clusters and fields and hence is unlikely to be affected by the foreground MW contamination. 

In the following subsections, we discuss the important results we obtained from the modeling. We compare the estimated kinematic parameters with the previous studies, analyze the rotation curve of the LMC for the control samples, trace the spatial variation of residual PM across the LMC, analyze the kinematic outliers from the model, and comment on the kinematic variation in parameters based on cluster age.

\subsection{Comparison of kinematic properties with previous studies}
The estimated kinematic parameters of the LMC are compared with those from previous studies, mainly in which either the authors estimated the kinematic centers and COM PM values or adopted them. First we compare the estimations of ($\alpha_0$, $\delta_0$), ($\mu_{W,com}$, $\mu_{N,com}$), $i$, and $\Theta$ from studies by N22, C22, G21, W20, and \citet[hereafter V14]{vande2014ApJ...781..121V}. The parameters estimated are tabulated in Table \ref{tab_comparison_parameters} along with estimations from this study for comparison.

\begin{table}
    \centering
    \caption{Comparison of the estimated kinematic parameters with previous studies. The first two rows list the estimations from this study.}
    \label{tab_comparison_parameters}
    \renewcommand{\arraystretch}{1.5}
    \begin{tabular}{c@{\hspace{0.85em}}c@{\hspace{0.85em}}c@{\hspace{0.85em}}c@{\hspace{0.85em}}c@{\hspace{0.85em}}c@{\hspace{0.85em}}c@{\hspace{0.85em}}l}
        \hline
        \hline
        \textbf{$\alpha_0$} & \textbf{$\delta_0$} & \textbf{$\mu_{W,\text{com}}$} & \textbf{$\mu_{N,\text{com}}$} & \textbf{$i$} & \textbf{$\Theta$}& Reference \\
        (deg)& (deg) & (deg) & (deg) & (deg) & (deg) &  \\
        \hline

        81.69$^{\hspace{0.01cm}\scaleto{+0.12}{4.7pt}}_{\hspace{0.01cm}\scaleto{-0.12}{4.7pt}}$ & -69.59$^{\hspace{0.01cm}\scaleto{+0.03}{4.7pt}}_{\hspace{0.01cm}\scaleto{-0.03}{4.7pt}}$ &-1.849$^{\scaleto{+0.003}{4.7pt}}_{\scaleto{-0.003}{4.7pt}}$ &0.397 $^{\scaleto{+0.007}{4.7pt}}_{\scaleto{-0.007}{4.7pt}}$ & 31.39$^{\hspace{0.01cm}\scaleto{+0.73}{4.7pt}}_{\hspace{0.01cm}\scaleto{-0.74}{4.7pt}}$ & 122.22$^{\hspace{0.01cm}\scaleto{+1.48}{4.7pt}}_{\hspace{0.01cm}\scaleto{-1.51}{4.7pt}}$ &Clusters \\

        80.57$^{\hspace{0.01cm}\scaleto{+0.07}{4.7pt}}_{\hspace{0.01cm}\scaleto{-0.07}{4.7pt}}$ & -69.34$^{\hspace{0.01cm}\scaleto{+0.02}{4.7pt}}_{\hspace{0.01cm}\scaleto{-0.02}{4.7pt}}$ &-1.851$^{\scaleto{+0.002}{4.7pt}}_{\scaleto{-0.002}{4.7pt}}$ &0.344 $^{\scaleto{+0.004}{4.7pt}}_{\scaleto{-0.004}{4.7pt}}$ & 30.81$^{\hspace{0.01cm}\scaleto{+0.38}{4.7pt}}_{\hspace{0.01cm}\scaleto{-0.40}{4.7pt}}$ & 128.02$^{\hspace{0.01cm}\scaleto{+0.79}{4.7pt}}_{\hspace{0.01cm}\scaleto{-0.81}{4.7pt}}$ &Field \\
        
        $79.95^{\scaleto{+0.22}{4.7pt}}_{\scaleto{-0.23}{4.7pt}}$ & $-69.31^{\scaleto{+0.12}{4.7pt}}_{\scaleto{-0.11}{4.7pt}}$ & $-1.867^{\scaleto{+0.008}{4.7pt}}_{\scaleto{-0.008}{4.7pt}}$ & $0.314^{\scaleto{+0.014}{4.7pt}}_{\scaleto{-0.014}{4.7pt}}$ & $33.5^{\scaleto{+1.2}{4.7pt}}_{\scaleto{-1.3}{4.7pt}}$ & $129.8^{\scaleto{+1.9}{4.7pt}}_{\scaleto{-1.9}{4.7pt}}$ & N22 \footnote{Estimated using VMC data}\\

         $81.07$ & $-69.41$ & $-1.862^{\scaleto{+0.002}{4.7pt}}_{\scaleto{-0.002}{4.7pt}}$ & $0.383^{\scaleto{+0.002}{4.7pt}}_{\scaleto{-0.002}{4.7pt}}$ & $28.7^{\scaleto{+1.4}{4.7pt}}_{\scaleto{-1.5}{4.7pt}}$ & $126.0^{\scaleto{+2.5}{4.7pt}}_{\scaleto{-2.6}{4.7pt}}$ & N22 \footnote{Estimated with fixed ($\alpha_0$, $\delta_0$) based on G21, using VMC data}\\
        
         $80.443$ & $-69.272$ & $-1.859$ & $0.375$ & $23.396^{\scaleto{+0.493}{4.7pt}}_{\scaleto{-0.501}{4.7pt}}$ & $138.856^{\scaleto{+1.360}{4.7pt}}_{\scaleto{-1.370}{4.7pt}}$ & C22 \footnote{Estimated with fixed ($\alpha_0$, $\delta_0$) and COM PM, using RC population with \textit{Gaia EDR3}} \\
         
         $81.07$ & $-69.41$ & $-1.847$ & $0.371$ & $33.28$ & $130.97$ & G21 \footnote{Estimated with \textit{Gaia} EDR3} \\

        $81.28$ & $-69.78$ & $-1.858$ & $0.385$ & $34.08$ & $129.92$ & G21 \footnote{Estimated with fixed ($\alpha_0$, $\delta_0$), using \textit{Gaia} EDR3} \\
         
         $80.90\pm0.29$ & $-68.74\pm0.12$ & $-1.878\pm0.007$ & $0.293\pm0.018$ & $25.6\pm1.1$ & $135.6\pm3.3$ & W20 \footnote{Estimated using carbon stars, with SkyMapper DR1}\\

         $81.23\pm0.04$ & $-69.00\pm0.02$ & $-1.824\pm0.001$ & $0.355\pm0.002$ & $26.1\pm0.1$ & $134.1\pm0.4$ & W20 \footnote{Estimated using RGB stars, with SkyMapper DR1} \\
         
         $80.98\pm0.07$ & $-69.69\pm0.02$ & $-1.860\pm0.002$ & $0.359\pm0.004$ & $29.4\pm0.4$ & $152.0\pm1.0$ & W20 \footnote{Estimated using Young stars, with SkyMapper DR1} \\

        $78.76\pm0.52$ & $-69.19\pm0.25$ & $-1.910\pm0.020$ & $0.229\pm0.047$ & $39.6\pm4.5$ & $147.4\pm10.0$ & V14 \footnote{Estimated with the third epoch of Hubble Space Telescope (HST) data, using PM.}\\

        $79.88\pm0.83$ & $-69.59\pm0.25$ & $-1.895\pm0.024$ & $0.287\pm0.054$ & $34.0\pm7.0$ & $139.1\pm4.1$ & V14 \footnote{Estimated with the third epoch of HST data, using PM + v$_{LOS}$ of old stars.}\\

        $80.05\pm0.34$ & $-69.30\pm0.12$ & $-1.891\pm0.018$ & $0.328\pm0.025$ & $26.2\pm5.9$ & $154.5\pm2.1$ & V14 \footnote{Estimated with the third epoch of HST data, using PM + v$_{LOS}$ of young stars.} \\

         \hline
         \hline
    \end{tabular}
\end{table}
\begin{figure*}
    \centering
       \includegraphics[width=1\linewidth]{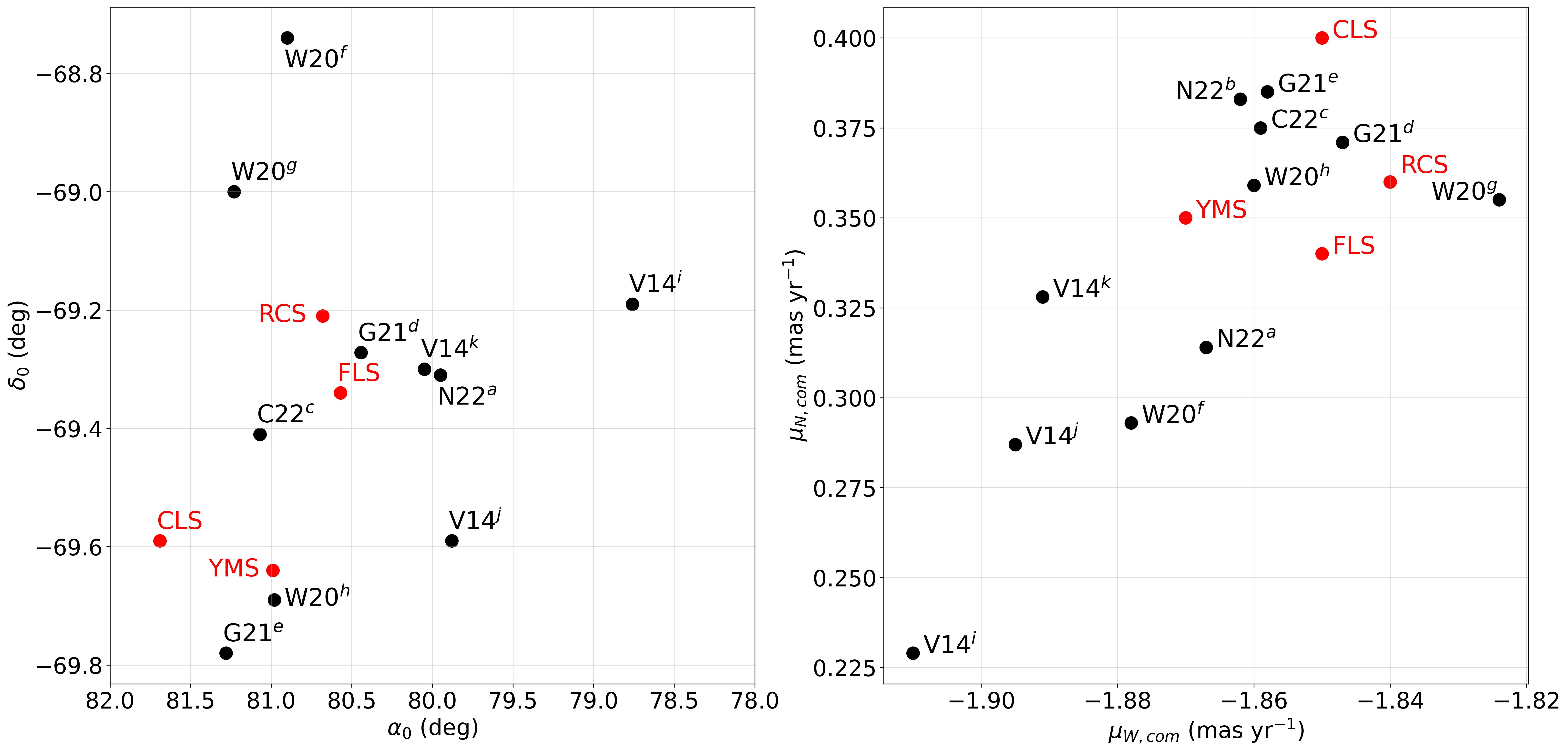}
       \caption{The parameter space of the estimated ($\alpha_0$, $\delta_0$) and ($\mu_{W,com}$, $\mu_{N,com}$) are compared with the reference studies (Table \ref{tab_comparison_parameters}) are provided in panels (a) and (b), respectively. The corresponding estimated parameters for the clusters (CLS), field (FLS), young MS stars (YMS), and RC stars (RCS) are marked with red dots along with labels. The reference studies are marked with black dots along with labels. }
       \label{fig:Reference_studies} 
\end{figure*}
Figure \ref{fig:Reference_studies} shows the distribution of  ($\alpha_0$, $\delta_0$) and ($\mu_{W,com}$, $\mu_{N,com}$)  estimations from the literature studies as mentioned above, along with the estimations from this study. We can notice a significant variation in the values of ($\alpha_0$, $\delta_0$) from various studies. Our estimations (shown in red) are located well within the range of other estimations, except the $\alpha_0$ for the cluster population. Most clusters ($\sim$ 75\%) used for the modeling in this study are of ages younger than 1 Gyr (D24). Due to this, the estimated values of ($\alpha_0$, $\delta_0$) for clusters are closer to the estimation by W20 using the young stellar population. Meanwhile, the estimation from the young MS population is much closer to the W20 estimations. The field population and RC population are closely in agreement with the recent studies by G21, V14, N22, and C22. The offset of ($\alpha_0$, $\delta_0$) in the older and younger stellar population was previously noted by W20.

The $\mu_{W, com}$ and $\mu_{N, com}$ values for the clusters and field population also align with the recent studies by N22, C22, G21, and W20 in a similar manner. Notably, the clusters show a relatively larger PM in the North direction when compared to the field. However, the kinematic centers and COM PM are distinct for the clusters and field population. 

The values of $i$ and $\Theta$ estimated in this study are compared with the previous studies (see Table \ref{tab_comparison_parameters}) and are more or less in agreement. However, the kinematic parameters such as $i$ and $\Theta$ are known to vary with respect to the coverage of the LMC (N22). The variations, as seen from the table, also contribute to the coverage. 
In this study, for the first time, we studied the variation of kinematic parameters as a function of the spatial coverage of the LMC using star clusters, as described in the subsection \ref{subsec_coverage}. The inner regions of the LMC show a larger inclination with respect to outer regions, which is already noted in the study by \cite{saroo2022A&A...666A.103S} using RC population.

The following section discusses the rotational parameters ($v_0$, $R_0$) and rotation curves of the LMC for various populations.

\subsection{Rotation of the LMC}\label{km_disc_rotation}

\begin{figure*}
    \centering
       \includegraphics[width=1\linewidth]{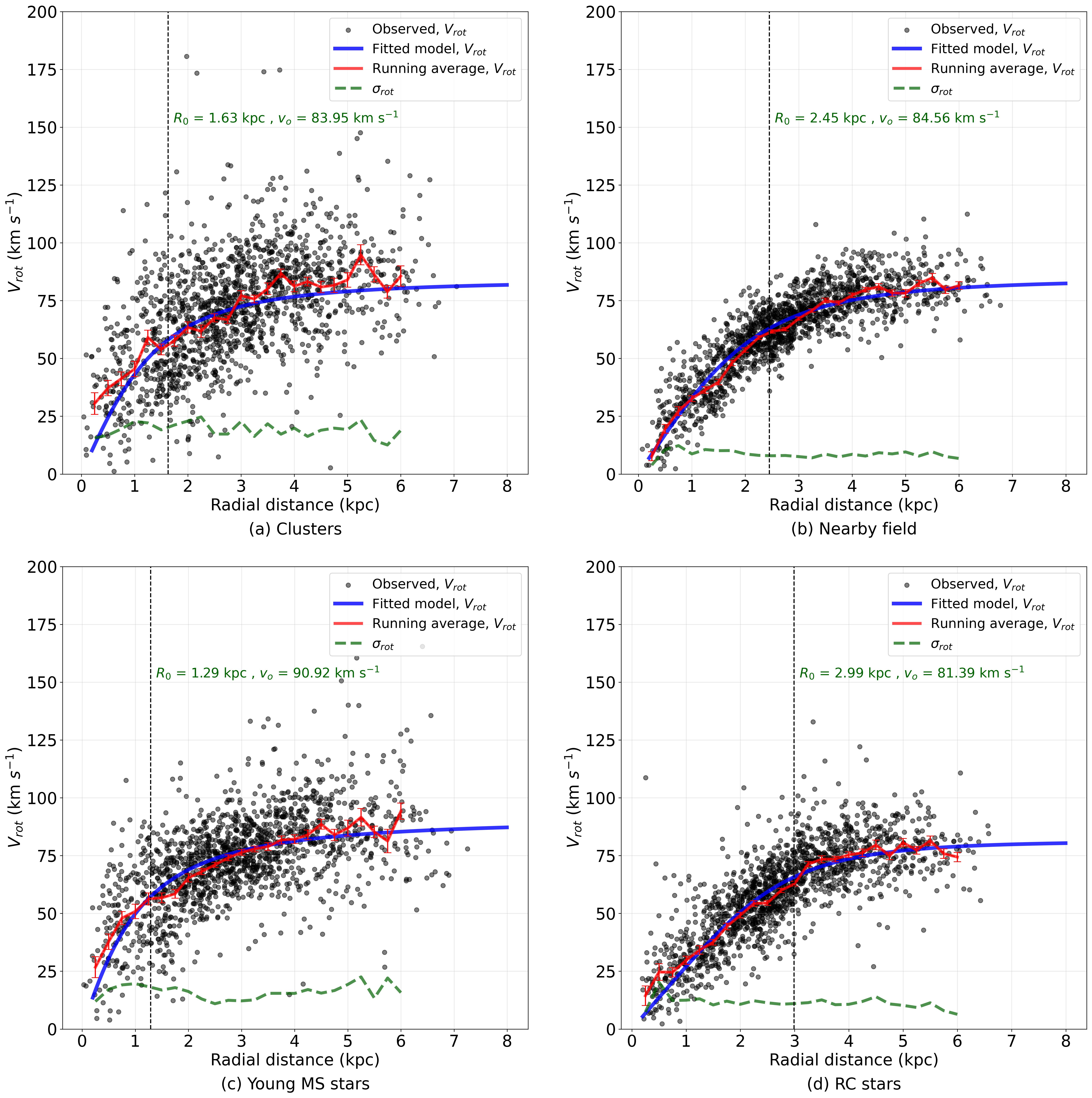}
       \caption{The rotation curve of the LMC based on the parameter estimation for the control sample (Clusters, Nearby field, young MS stars, and RC stars) are plotted in panels (a) to (d). The magnitude of the rotational velocity in the LMC plane (V$_{rot}$) is shown with black dots in each panel. The red curve represents the running average over the observed V$_{rot}$ with a bin size of 0.25 kpc, and the blue curve shows the best-fitting model depending on the rotational parameters ($v_0$, $R_0$, $\eta$) estimated for each data set (see Table \ref{tab3_paraemters}).}
       \label{fig: Rotation curves} 
\end{figure*}

The rotation curves for the clusters, nearby fields, young MS stars, and RC stars from the control sample (as mentioned in subsection \ref{km_control}) in the LMC plane, after deprojection are shown in Figure \ref{fig: Rotation curves} (a. Clusters, b. Nearby field, c. young MS stars, d. RC stars). The running average for the data points is made with a bin resolution of 0.25 kpc, and shown in red. The fitted model is shown with the blue curve. The rotational velocity amplitude ($v_0$) does not show a significant variation across the clusters and field population. The estimated $v_0$ for clusters, fields, and RC stars are consistent with the estimates by W20. The young MS stars in our study show a slightly larger $v_0$ ($\sim$ 91 km s$^{-1}$) when compared to the RC stars. However, it is consistent with the observed $v_0$($\sim$ 90 km s$^{-1}$) for the younger population studied by N22.  

In the cluster age-dependent kinematic model (subsection \ref{km_age}), the variation of $R_0$ for cluster age groups is provided. The age group C$_{AG-3}$ (log(age) = 8.0 - 8.65) and young MS stars from the control population show reduced values of $R_0$ at 1.34$^{\hspace{0.01cm}\scaleto{+0.17}{4.7pt}}_{\hspace{0.01cm}\scaleto{-0.16}{4.7pt}}$ kpc and 1.29$^{\hspace{0.01cm}\scaleto{+0.08}{4.7pt}}_{\hspace{0.01cm}\scaleto{-0.08}{4.7pt}}$ kpc, respectively. This is suggestive of a steeper rise of the rotational velocity when compared to the older population. 
The decreased $R_0$ for these age groups may imply a redistribution of mass, leading to a higher mass density in the central regions. Also, in the youngest group C$_{AG-4}$ (log(age) = 6.55 - 8.0), the cluster density in the galaxy shifts towards the North-East regions (D24). 
In C$_{AG-4}$, there is a shift in the value of $R_0$ to a slightly higher value of 1.98$^{\hspace{0.01cm}\scaleto{+0.17}{4.7pt}}_{\hspace{0.01cm}\scaleto{-0.18}{4.7pt}}$ kpc. The field population near clusters is dominated by the older population, hence the $R_0$ value remains high. 

\begin{figure*}
    \centering
       \includegraphics[width=1\linewidth]{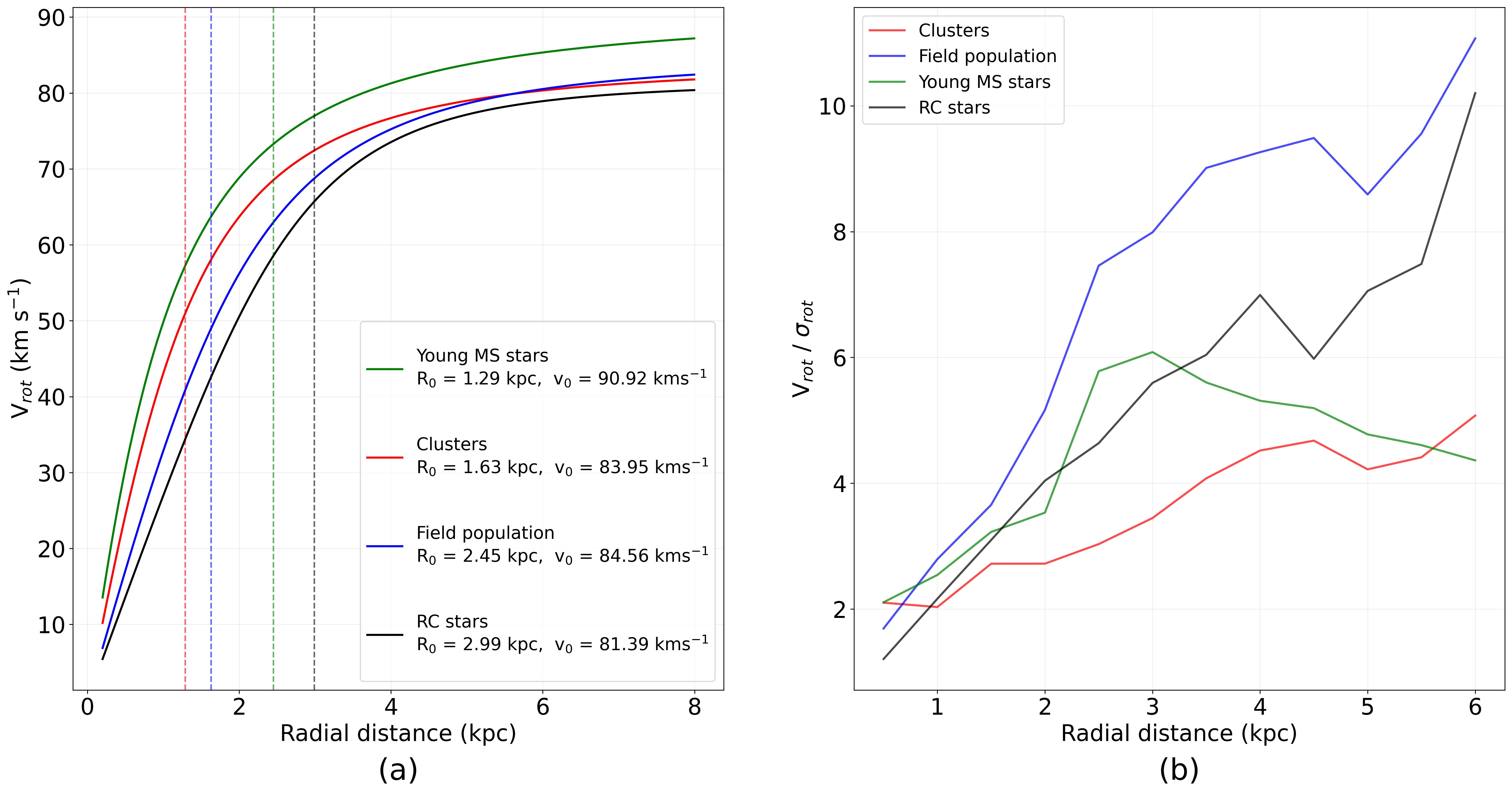}
       \caption{Variataion of $R_0$ and $V_{rot}$/$\sigma_{rot}$ for the control population (Clusters, Nearby field, young MS stars, and RC stars). (a) Modeled velocity profiles showing the variation in $R_0$ across the control sample; (b) Variation of $V_{rot}$/$\sigma_{rot}$ across the control sample, estimated using a bin size of 0.4 kpc in radii. }
       \label{fig: sec_5_2} 
\end{figure*}

Panel (a) of Figure \ref{fig: sec_5_2} shows all the four model velocity profiles fitted to the control population provided in Figure \ref{fig: Rotation curves}. The $v_0$ for the control population is almost identical ($\sim$ 81 to 84 kms$^{-1}$), whereas the value for the young MS stars remains slightly higher ($\sim$ 91 kms$^{-1}$). It is clear from this figure that the clusters and young MS stars show a steeper rise of the V$_{rot}$ suggesting an increased mass density in the inner regions. A similar trend of small $R_0$ in the case of the younger population and larger $R_0$ in the case of the older population is noted in the study by W20. The variation in the value of  $R_0$ with respect to population may be due to the effect of the evolution of the bar and its activity with time, resulting in the redistribution of mass in the inner regions.

Notably, the dispersion of the rotation velocity profile resides at $\sim$ 23 km s$^{-1}$ and $\sim$ 11 km s$^{-1}$ up to 6 kpc in the case of clusters and field population, respectively. In the case of the young MS population, there is a larger $\sigma_{rot}$ of $\sim$ 22 km s$^{-1}$ up to 2 kpc, then a slight decline and further rises after 4 kpc. Meanwhile, the RC population has a $\sigma_{rot}$ of $\sim$ 20 km s$^{-1}$ within 1 kpc and slowly declines and levels off at $\sim$ 12 km s$^{-1}$ afterward. The larger dispersion values in the central 2 kpc might be due to the non-circular motions due to the bar. 

Panel (b) of Figure \ref{fig: sec_5_2} shows the radial variation of the  V$_{rot}$/$\sigma_{rot}$ ratio for the four control population. As all of the population have similar V$_{rot}$ (except slightly higher value for the young MS stars), the higher values observed for the field stars suggest a relatively low $\sigma_{rot}$. The RC population has a similar profile, though with a slightly lower ratio (suggesting a relatively large $\sigma_{rot}$). The young MS stars have a more or less similar profile till $\sim$ 3 kpc, and we note a lower value of the ratio beyond this radius (suggesting a relatively large $\sigma_{rot}$). The cluster has a shallow profile and lowest value for the ratio, suggestive of the highest value of $\sigma_{rot}$ at all radii.
Overall, the low value of  $\sigma_{rot}$ the field population points to the minimal disk heating for this population. However, the ratios suggest that young clusters and young stars have relatively large velocity dispersion, which is suggestive of relatively recent perturbation(s) in the LMC disk. This may point to the heating of the gas in the disk (resulting in the formation of stars with similar kinematics), with minimal heating of the stellar disk. The heating may be due to internal perturbations, such as the bar and spiral arms, or external perturbation, which is the interaction with the SMC. 

The rotational velocity maps shown in Figure \ref{fig:Rot_spatial_4_pop} closely match with the velocity maps from G21. Additionally, the variation in slope between evolved and younger populations observed in Figure \ref{fig: sec_5_2} is consistent with G21. However, there is a discrepancy in the rotation velocity dispersion profiles when compared to G21. 
\subsection{Spatial variation of residual PM: Clusters and Field}\label{km_disc_discheating}
\begin{figure*}
    \centering
       \includegraphics[width=0.55\linewidth]{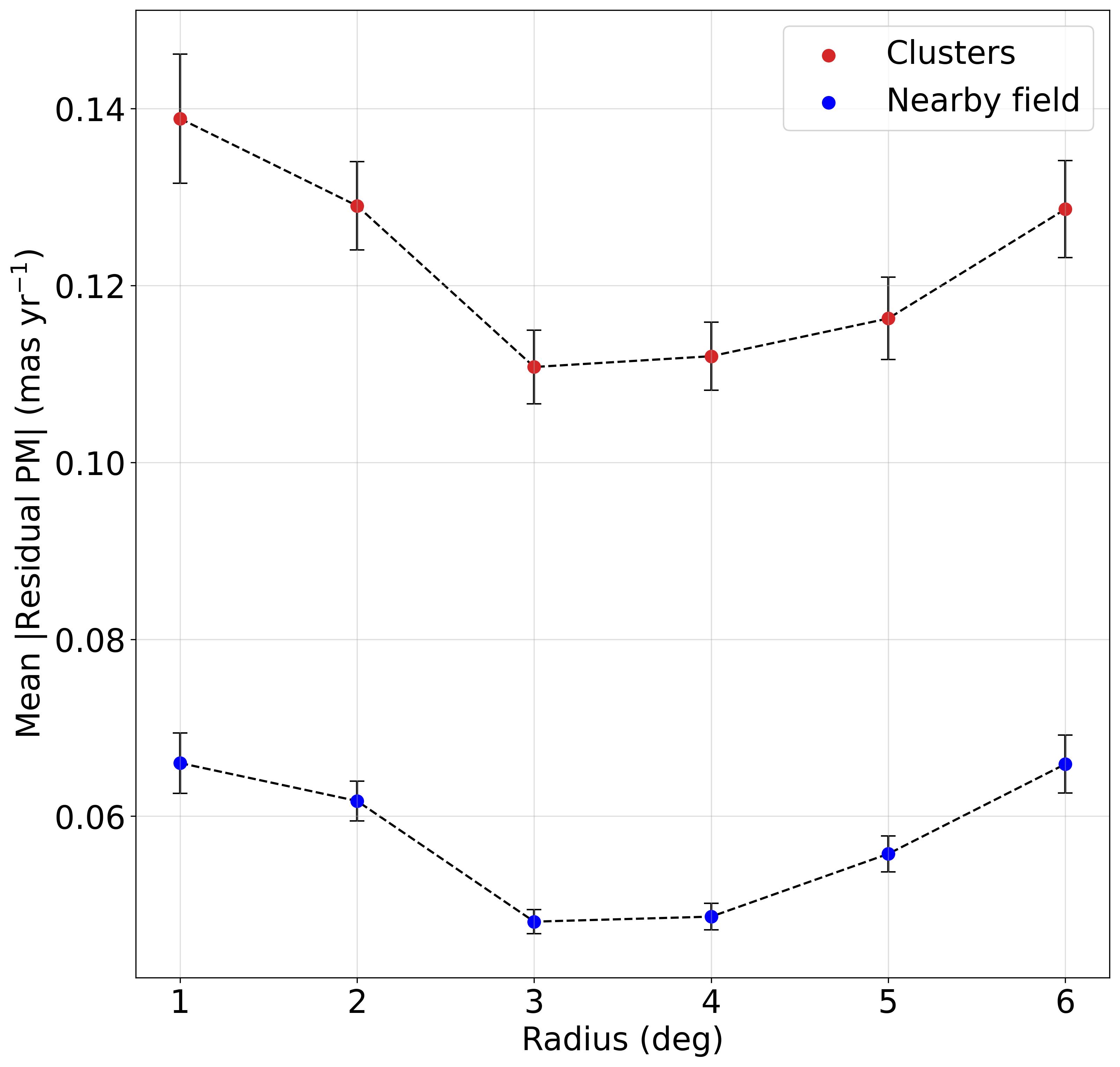}
       \caption{The radial variation of mean $|$Residual PM$|$ in the sky plane. Annular regions with a bin size of 1$^\circ$ are used. Mean $|$Residual PM$|$ of clusters and field population are marked with red and blue dots, respectively. }
       \label{fig:respmquadrant_plots} 
\end{figure*}

In order to trace regions with large variations in the residual PM, 
the mean $|$Residual PM$|$ is traced for different radii (1 to 6$^\circ$ with a bin size of 1$^\circ$ using annular regions) as shown in Figure \ref{fig:respmquadrant_plots}. We observe a similar trend in radial variation of mean $|$Residual PM$|$ for both the clusters and field population, though the values are significantly different. For inner radii less than 2$^\circ$, the mean residual PM of the LMC is larger for both populations, likely to be due to the presence of the bar. Beyond 2$^\circ$, we note a decline in mean residual PM, and it increases after 4$^\circ$. This could be due to the presence of spiral arms in the galaxy, where dense star formation is noted. However, there is a clear signature of larger residual PM for clusters when compared to the field population. 

The residual PM RMS profiles in Figure \ref{fig:respm_plots} can be compared with the study by C22 based on the numerical simulations by B12 with a population older than 1 Gyr. Figure 8 in C22 has provided the predicted $log \text{RMS}_{|\text{Residual PM}|}$ as a function of the impact parameter (P in kpc) and impact timing (T in Myr) of the recent LMC-SMC interaction. We adopted the value of T as 149 Myr, based on the significant CF peak in the SMC as observed in Figure 8b of D24 to estimate the likely value of P. We note the assumed value of T is also in good agreement with the estimate by \cite{zivick2018ApJ...864...55Z}, who suggests a recent direct collision between the LMC and the SMC at 147 Myr ago. The field $log \text{RMS}_{|\text{Residual PM}|}$ of $-$1.611 and $-$1.219 from the primary model and control sample support the Future 40 and Future 60 models provided in C22. We estimate that the impact parameter, P, must reside at less than 3 kpc, and most likely match with the impact parameter of 2 kpc as per B12 simulations. However, the RC stars in the control sample show a larger residual (0.08 mas yr$^{-1}$) compared to the estimates of C22 (0.058 mas yr$^{-1}$). We note that our model involves treating the COM PM and kinematic centers as free parameters, unlike C22.  
Hence a direct comparison of our results with C22 may be unrealistic.

The clusters and young stars show larger disc heating, with $log \text{RMS}_{|\text{Residual PM}|}$ of -0.853 and -0.957. This suggests that the young population is more perturbed, unlike the RC stars and other field stars. Also, the clusters, irrespective of age, show larger residual PM. The kinematic deviant clusters are analyzed in the next section.

\subsection{Kinematically deviant clusters: Tracing the regions of larger residual PM}
\begin{figure*}
    \centering
       \includegraphics[width=1\linewidth]{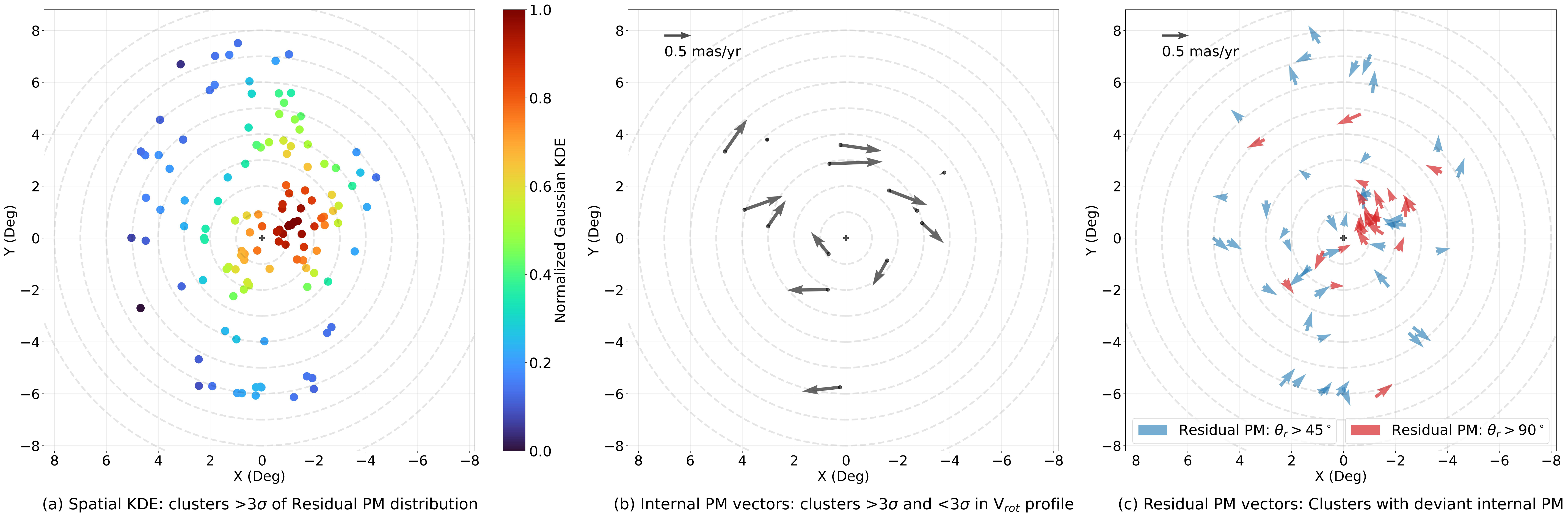}
       \caption{The clusters of larger residual PM are plotted here. (a) Spatial 2D Gaussian KDE of cluster locations with $>3\sigma$ of $|$Residual PM$|$; (b) Internal PM vectors of clusters $>3\sigma$ and $<3\sigma$ in V$_{rot}$ profile of Figure \ref{fig: Rotation curves}(a); (c) Residual PM vectors of clusters with model and observed internal PM vectors differing by  $\theta_r>$45$^\circ$ and $\theta_r>$90$^\circ$}.
       \label{fig:Res_PM_deviations} 
\end{figure*}

The residual PM distribution of clusters, as shown in Figure \ref{fig:respm_plots}a suggests that some clusters have a large residual PM. We traced clusters with significantly large residuals ($>$3$\sigma$ of $|$Residual PM$|$) and identified the region with the highest density of such clusters, as illustrated in Figure \ref{fig:Res_PM_deviations}(a) with the 2D Gaussian KDE. We find a significant density of clusters in the North-West at $\sim$ 1-2$^\circ$ with large residual PM. The bar region extending from North-West to South-East of the LMC shows relatively large residuals in PM.  

We note several outlier clusters in the observed v$_{rot}$ profile (Figure \ref{fig: Rotation curves}) of the LMC. The clusters falling outside the 3$\sigma$ margin from the running average of the v$_{rot}$ profile are treated as outliers. Their locations are traced, and their internal PM is plotted in Figure \ref{fig:Res_PM_deviations}b. 

The residual PM vectors are plotted in Figure \ref{fig:Res_PM_deviations}c for clusters with the relative angular difference ($\theta_r$) between the observed and modeled internal motion greater than 45$^\circ$ and 90$^\circ$.  The red vectors indicate clusters with $\theta_r >$ 90$^\circ$, suggestive of counter-rotation. These are dominantly seen in the North-West of the galaxy. The deviant clusters also trace the non-circular motion of the bar. 

The North-West region with the largest residual PM is located between a radial distance of 1-2$^\circ$. The same location also shows the presence of counter-rotating clusters. This region has PM deviation both in angle (panel c)  and in value (panel a), pointing to the region experiencing an event of perturbation. As this region is located at an impact distance similar to that estimated by C22 and in this study, we speculate that this may be the area that experienced the largest impact due to the recent LMC-SMC collision. We also note that there could also be other reasons, such as perturbations that could occur at the ends of the bar, but a similar disturbance is not seen at the eastern end of the bar. Hence bar perturbation may not be the reason for this disturbance.

\subsection{Kinematic variation of parameters in the younger clusters}\label{age_young_disc}

The kinematic model of the LMC for different age groups is studied as in subsection \ref{km_age}. We note the kinematic parameters of the LMC, mainly the $i$, $\Theta$, and $R_0$, show variation in the younger age group C$_{AG-3}$ (100-440 Myr). To further investigate the change in kinematic parameters, we split the $C_{AG-3}$ into two groups (one from 100 to 200 Myr and the other from 200 to 440 Myr) and performed the kinematic modeling. The variation of $i$, $\Theta$, and $R_0$ are shown in Figure \ref{fig:Age_dependence_kineamtics}.

The age group from 100 to 200 Myr shows the largest inclination ($\sim$ 39$^{\circ}$) compared to other age groups. Meanwhile, we note the smallest $\Theta$ value ($\sim$ 105$^{\circ}$) in this age range when compared to other age ranges. The cluster density shifted towards central regions of the LMC in the age group C$_{AG-3}$ and towards North-East in C$_{AG-4}$ (D24). We note this effect in the variation of $R_0$ to the inner radii of the galaxy, as shown in Figure \ref{fig:Age_dependence_kineamtics}c. The value of $R_0$ reaches 1.1 kpc for clusters of age between 100 to 200 Myr. A similar effect is seen in the nearby young population, where the value reaches 0.5 kpc. Meanwhile, the field population shows smaller $i$ and lesser variation in $R_0$ compared to other age groups. The kinematic model of cluster groups based on richness also shows this trend of variation in $R_0$. The younger and richer clusters with more than 60 members are located more toward the central regions of the galaxy as well. We note $R_0$ smaller than 1 kpc at this cluster group. In summary, we note that the $i$, $\Theta$, and $R_0$ values for clusters and young MS stars in the age range 100-200 Myr are distinct from the other groups. A possible reason for this deviation is discussed in subsection \ref{disc:lmc_smc_int}. 

\subsection{Kinematic signatures of the LMC bar}\label{disc:bar}

The spatial PM plots of clusters and field population, as shown in Figure \ref{fig:PM_plots}, do not bring out the bar feature explicitly. The residual PM plots, as shown in Figure \ref{fig:respm_plots}, on the other hand, show the bar feature, though more prominent in the case of clusters. The radial gradient of the residual PM (Figure \ref{fig:respmquadrant_plots}, panel a) shows a large gradient in the bar region, pointing to residuals resulting from kinematics related to the bar. The residual PM plots shown in Figure \ref{fig:Res_PM_deviations} (panels a and c) show the possible presence of non-circular motions in the bar region. 
A significant reduction in the value of $R_0$ for clusters and young MS stars with respect to RC stars (see Figure \ref{fig: sec_5_2}a) is probably related to the mass distribution within the inner 2 kpc, where the bar is present. The fact that the younger population reaches the maximum value of V$_{rot}$ in a shorter radius when compared to the older population points to a redistribution of mass, most likely due to the evolution of the bar. 

In summary, this study has traced three kinematic signatures of the LMC bar, relatively large residual PM in the inner 2$^\circ$ radius for both clusters and field population, the presence of non-circular motion among star clusters, and a decrease of $R_0$ as a result of the possible evolution of the bar. Recently, J23 also found that the dynamics of the inner disk are dominated by the bar.

\subsection{Effect of the recent LMC-SMC interaction on the LMC disk}\label{disc:lmc_smc_int}

The panels (a) and (c) of Figure \ref{fig:Res_PM_deviations}, show that though there are kinematically deviant clusters spread through the LMC disk, there is a specific location where the deviation is maximum. Panel (c) indicates that the same location also shows the presence of cluster motion in all directions, suggesting a possible specific external disturbance. It is therefore important to identify the source of the disturbance identified in this study. 

\begin{figure*}
    \centering
       \includegraphics[width=1\linewidth]{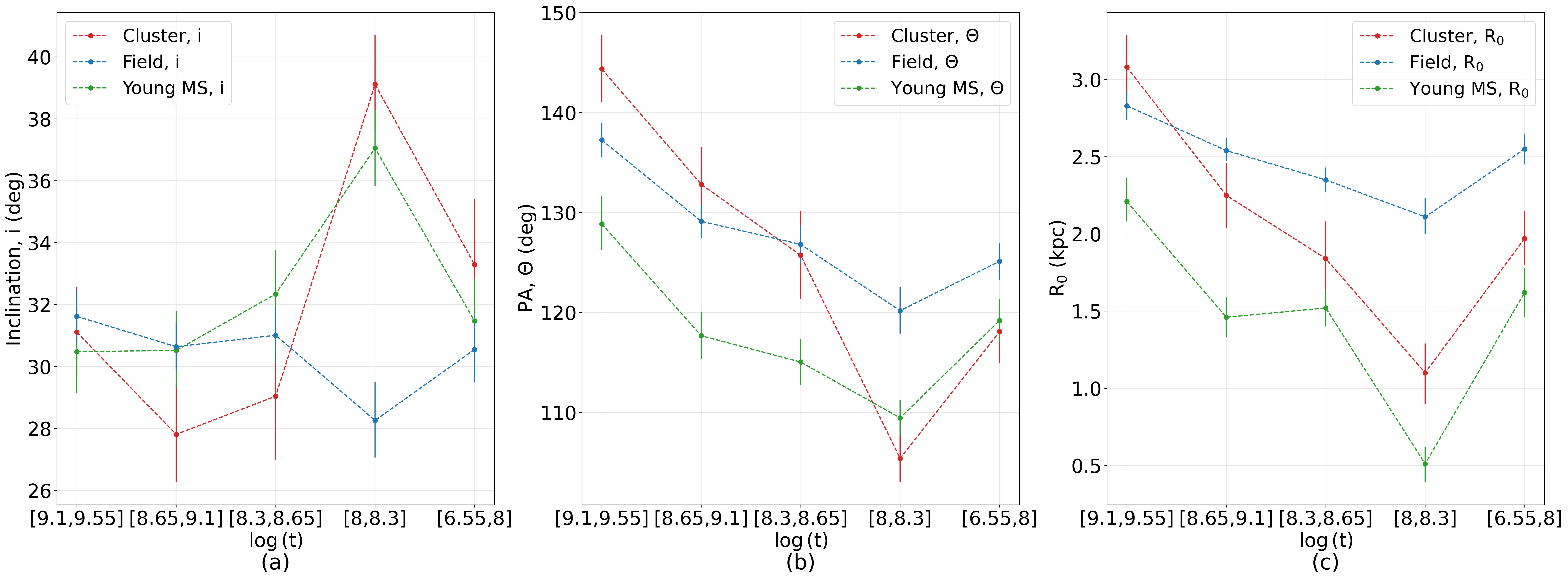}
       \caption{The variation of $i$, $\Theta$, and $R_0$ with respect to the cluster age groups (subsection \ref{age_young_disc}) are shown here. Red, Blue, and Green markers are used for clusters, fields, and young MS stars, respectively. (a) Variation of $i$; (b) Variation of $\Theta$ and (c) Variation of $R_0$ are shown for 5 age groups. The age groups are valid only for clusters (Section \ref{sec_2}). The field and the young MS stars are age-wise heterogeneous but located near the clusters within each age group.}
       \label{fig:Age_dependence_kineamtics} 
\end{figure*}

It is possible that this disturbance is caused by the LMC-SMC collision at $\sim$ 150 Myr ago as the radial distance of this region from the center also matches the value of the impact factor of the LMC-SMC collision as derived by C22. The age-dating of the disturbance, as shown in Figure \ref{fig:Age_dependence_kineamtics} suggests a significant shift in inclination $i$, PA, and $R_0$ for clusters in the age range 100-200 Myr. The kinematic parameters of clusters younger than this period are more similar to the overall disk properties. This is an indication that the clusters formed during this period have significantly different kinematic parameters. We speculate that this is the first evidence of direct collision in the LMC disk and the spatial and temporal kinematic disturbance, matching with the predicted time and location of the LMC-SMC collision. We also speculate that the interaction mainly affected the gas, and the clusters and stars born from the disturbed gas bear the signature of the perturbation. 

\section{Summary}\label{sec_6}
We summarise the results and conclusions from this comprehensive study of the LMC disk kinematics using \textit{Gaia} DR3 data.
\begin{enumerate}
    \item We performed the kinematic model of the LMC that corresponds to the observed median PM using 1705 star clusters and field regions. This is a comprehensive study with a total of 48 data sets analyzed to create 48 models to study the dependence of the kinematic model on various parameters.
    \item The model PM for the tracers in the data sets is formulated based on the equations outlined by V02.  The model parameters estimated include inclination of the LMC disc ($i$), the position angle of the line of nodes  ($\Theta$), dynamic centers ($\alpha_0$, $\delta_0$), the amplitude of the tangential velocity of the LMC's center of mass ($v_t$), the tangential angle made by $v_t$ ($\theta_t$), scale radius ($R_0$), optimization factor ($\eta$), and the amplitude of the rotational velocity ($v_0$). We assumed a fixed distance to the LMC center and a line-of-sight velocity of the COM, $v_{sys}$. The fitting of the parameters was performed by an MCMC technique, and model parameters were estimated for all datasets listed in Table \ref{tab1}.
    \item This is the first 2D kinematic model of the LMC employing clusters and neighboring field regions with \textit{Gaia} DR3 data. The data coverage of the LMC considered in this study is within $\sim$ 7$^\circ$ from the LMC center. The parameters estimated in this study show good agreement with estimations in the literature when the comparison is made between similar populations.
\item There is no significant difference between the observed COM PM between cluster and field. We note an offset of 28$\pm$8 \text{arcmin} between the dynamic centers ($\alpha_0$, $\delta_0$) of the cluster and field. Estimated $\Theta$ values of 122$^\circ$.22$^{\hspace{0.01cm}\scaleto{+1.48}{4.7pt}}_{\hspace{0.04cm}\scaleto{-1.51}{4.7pt}}$ for clusters and 128$^{\circ}$.02$^{\hspace{0.01cm}\scaleto{+0.79}{4.7pt}}_{\hspace{0.04cm}\scaleto{-0.81}{4.7pt}}$ for field regions point to an offset of offset of 5$^\circ$.8$\pm$1.7 between them, while the inclination, $i$, remains almost similar. The modeled rotational parameters, $R_0$ and $\eta_0$ appear to be larger for field regions compared to clusters, while the $v_0$ remains almost similar.
\item We also estimated the kinematic model parameters for two control populations, RC stars and young MS stars. We find that the younger population tends to show a southern and eastward dynamic center and a relatively smaller value of $R_0$. We also note a varying value of $\Theta$ across the control population. 
\item This study establishes that the kinematic model of the LMC disk varies with the age of the cluster population used for the estimation, in contrast to the surrounding field population. 
\item The estimated parameters show a significant radial dependence for both cluster and field population. The value of  $i$ steeply decreases from the inner to outer radii, from $\sim$ 50$^\circ$ to $\sim$ 32$^\circ$ for clusters, and $\sim$ 43$^\circ$ to $\sim$ 31$^\circ$ for field population. The value of $\Theta$ increases from inner to outer radii, from $\sim$ 108$^\circ$ to $\sim$ 122$^\circ$ for clusters, and $\sim$ 110$^\circ$ to $\sim$ 128$^\circ$ for field population. The rotational parameters, $R_0$ and $v_0$ after 5$^\circ$ show convergence to the values estimated from the full sample. 
\item Clusters show larger PM residuals when compared with the field population. The RMS distribution of the residual PM for clusters shows a broader profile, while the corresponding distribution for the field shows a narrower one.
\item The rotational velocity amplitude ($v_0$) for clusters, fields, and RC stars ($\sim$ 81 to 84 km s$^{-1}$) are consistent with the estimates by W20. The young MS stars in our study show a slightly larger $v_0$ ($\sim$ 91 km s$^{-1}$) when compared to the RC stars, as also noted by N22.  
\item The modeled rotational parameters, $R_0$ and $\eta_0$ appear to be larger for field population when compared to clusters, while the $v_0$ remains almost similar. We note a significant shift in $R_0$ of $\approx$ 1.7 kpc between young MS and RC population. The variation in the value of  $R_0$ with respect to population may be due to the redistribution of mass in the inner regions, and it may be due to the evolution of the bar and its activity over time.
\item The dispersion of the rotation velocity is found to be $\sim$ 23 km s$^{-1}$ for clusters, and is likely to have contributions from the bar and spiral arms. The value of $\sim$ 11 km s$^{-1}$ for field population is relatively low and suggests insignificant stellar disk heating. Young MS stars show a relatively large velocity dispersion, similar to clusters. 
\item The residual PM for the cluster and the field decreases from the center up to 3$^\circ$, then increases beyond 4$^\circ$. The increased value in the inner 2$^\circ$ region is likely to be the effect of the bar. We also detect evidence for non-circular motion in the bar region among the clusters.
\item Location of kinematically deviant clusters show an increased density in the North-West side. We also trace the presence of a number of counter-rotating clusters, mainly in the same region. 
\item We detect a specific kinematic perturbation between 2-3$^\circ$ from the center in the North-West direction. We also note a significant shift in the  $i$, $\Theta$, and $R_0$ of clusters in the age range of 100-200 Myr. This spatial and temporal disturbance matches with the impact factor and the time of the LMC-SMC collision as estimated by C22 and this study. Could this be the evidence for the recent LMC-SMC collision?
\end{enumerate}

% \appendix

\section{Acknowledgements}
We thank the referee for valuable suggestions that helped to improve the manuscript. Annapurni Subramaniam acknowledges support from the Science and Engineering Research Board (SERB) Power fellowship. This work made use of the optical data from the European Space Agency (ESA) mission \textit{Gaia} (\url{https://www.cosmos.esa.int/gaia}), which was processed by the \textit{Gaia} Data Processing and Analysis Consortium (DPAC,
\url{https://www.cosmos.esa.int/web/gaia/dpac/consortium}). The funding for the DPAC has been provided by national institutions, particularly the institutions participating in the \textit{Gaia} Multilateral Agreement. 

{\it Software:} ASTROPY \citep{astropy2013A&A...558A..33A}, SCIPY \citep{2020SciPy-NMeth}, MATPLOTLIB \citep{matplot2007CSE.....9...90H}, NUMPY \citep{numpy2020Natur.585..357H}, CORNER \citep{corner} 
\bibliography{sample631}{}
\bibliographystyle{aasjournal}

\end{document}